\documentclass[12pt]{article}
\usepackage{graphicx}
\usepackage{amsthm}
\usepackage{amsmath}
\usepackage{amsfonts}
\usepackage{graphicx}
\usepackage{fancyhdr}
\usepackage{cite}
\newcommand{\bm}[1]{\mbox{\boldmath$#1$}}

\newcommand{\OO}{\mbox{\boldmath $\Omega$}}

\newcommand{\av}{\mbox{\boldmath $a$}}

\newcommand{\ks}{\mbox{\boldmath $\xi$}}

\newcommand{\sbb}{\mbox{\boldmath $s$}}

\newcommand{\si}{\mbox{\boldmath $\sigma$}}

\newcommand{\bb}{\mbox{\boldmath $b$}}

\newcommand{\hQ}{\mbox{\boldmath$\hat{Q}$}}
\newcommand{\hp}{\mbox{\boldmath$\hat{p}$}}

\newcommand{\hA}{\mbox{\boldmath $\hat{A}$}}

\newcommand{\cb}{\mbox{\boldmath $\chi$}}

\newcommand\fr{\displaystyle\frac}
\newcommand{\htts}{\mbox{\boldmath$\hat{t}\kern1pt$}}

\begin{document}
\title{On EPR paradox, Bell's inequalities, and experiments that prove nothing}
\author{V.K.Ignatovich}
\maketitle
\begin{abstract}
This article shows that the there is no paradox. Violation of
Bell's inequalities should not be identified with a proof of non
locality in quantum mechanics. A number of past experiments is
reviewed, and it is concluded that the experimental results should
be re-evaluated. The results of the experiments with atomic
cascade are shown not to contradict the local realism. The article
points out flaws in the experiments with down-converted photons.
The experiments with neutron interferometer on measuring the
``contextuality'' and Bell-like inequalities are analyzed, and it
is shown that the experimental results can be explained without
such notions. Alternative experiment is proposed to prove the
validity of local realism.
\end{abstract}
\section{Introduction}
The EPR paradox~\cite{epr} was formulated in the following way:
with the common sense logic we can predict that a particle can
have precisely defined position $x$ and momentum $p$
simultaneously, however quantum mechanics (QM) forbids it because
operators $\hQ$ and $\hp$ corresponding to position and momentum
are not commuting, $[\hQ,\hp]\ne0$. Therefore QM is not complete.

The common sense logic was applied to the quantum mechanical state
of two particles, which interacted at some moment in the past and
then flew far apart. The distance becomes so large after the
separation, that there are no more interaction between the
particles.

The state of the two particles is described by a wave function
$\Psi(x_1,x_2)$, where $x_{1,2}$ are coordinates of the particles.
This wave function can be expanded over the eigen functions
$u_p(x_1)$ of the momentum operator $\hp$ of the first particle
\begin{equation}\label{en1}
\Psi(x_1,x_2)=\sum\limits_{p}\psi(p,x_2)u_p(x_1),
\end{equation}
where $\psi(x_2,p)$ are the expansion coefficients. Although
particles flew far apart, their wave function, according to QM, is
not a product of two independent functions. It is a superposition
of such products. This superposition is used to be called
``entangled'' state. EPR devised a model where coefficients
$\psi(p,x_2)$ are also the eigen functions of the momentum
operator $\hp$. Therefore, if measurement with the first particle
reveals it in the state $u_p(x_1)$ with a fixed momentum $p$, then
the second particle is also in the state with the fixed momentum
$p$. According to EPR logic, the second particle does not interact
with the first one, therefore it has momentum $p$ independently of
the measurements on the first particle.

Similar consideration with the expansion over the eigen functions
$v_q(x_1)$ of the position operator $Q$ shows that the measurement
of the position of the first particle immediately predicts the
position of the second one. Since the second particle does not
interact with the first one, position and momentum of the second
particle exist independently of the measurements on the first one.
However, this case is impossible because of the noncommutativity
of position and momentum operators. Therefore, we must admit that
either QM is incomplete and should be improved by introducing some
not yet known (hidden) parameters, or that QM is complete but
there is some kind of instant action at a distance that puts the
second particle into the state consistent with the apparatus
measuring the first particle.

Such action at a distance must exist only if one accepts
Copenhagen interpretation of QM. According to this interpretation,
performing a measurement in one part of space immediately leads to
a reduction of the entangled state to a single term everywhere in
space. We can avoid such drastic instantaneous reduction, if we
further follow the EPR logic. According to this logic, if the
measurement on the first particle reveals it in the state
$u_p(x_1)$, then the second particle is in the state
$\psi(p,x_2)$. This wave function belongs to the second particle
independently of the measurements performed on the first one. But,
if the second particle is described by $\psi(p,x_2)$, the first
particle can only be in the state $u_p(x_1)$. Therefore the state
of the two particles before any measurements is described by the
single product term $\psi(p,x_2)u_p(x_1)$ of the sum (\ref{en1}).
The measurement on the first particle reveals only which term is
present in the given event.

Bohm-Aharonov~\cite{ahb} reformulated the EPR paradox in terms of
the spin components of two particles. They considered a molecule
with zero angular momentum and its decay into two spin 1/2 atoms A
and B, that fly far apart in opposite directions. According
to~\cite{ahb}, if the atom A is transmitted through an analyzer
oriented along a vector $\av$, then the spin of the atom B is
directed with certainty along the vector $-\av$. This result looks
very natural from QM point of view, because we always can choose
direction $\av$ as a quantization axis. The spin of the particle A
is directed along or opposite $\av$, i.e. its projection on $\av$
is $\pm1/2$. Therefore, the projection of the spin of the particle
B is $\mp1/2$. In the following we can imagine that the projection
has plus sign when the particle is counted by the detector after
the analyzer, and minus sign when it is not (it can then be
counted by a different nearby detector, which detects particles
reflected by the analyzer).

J.S. Bell~\cite{bel} considered this situation as a predetermined
one. He decided to check whether it is really possible to
introduce such a hidden parameter $\lambda$, with which we can
exactly predict whether particle A will have projection +1/2 or
-1/2 on the direction $\av$, and therefore particle B will have
projections $\mp1/2$ on the same direction. Situation becomes more
difficult when the analyzer for the particle B is aligned along a
vector $\bb$ noncollinear to $\av$. Then the quantization axis for
the particle B is along $\bb$, projections of the spin on this
axis are the same $\pm1/2$ as on the axis $\av$, but now the signs
of the projections are not necessarily opposite to the signs of
the projections of the spin of the particle A on the axis $\av$.
Bell suggested that the hidden parameter $\lambda$ predetermined
the sign of the projections of the two particles on vectors $\av$
and $\bb$. He introduced well defined functions
$A(\av,\lambda)=\pm1$ and $B(\bb,\lambda)=\pm1$ such that for
every fixed $\lambda$ the result of measurement of two spin
components can be described by a predetermined product
$A(\av,\lambda)B(\bb,\lambda)=\pm1$. If a distribution of
$\lambda$ is described by a classical probability density
$P(\lambda)$, then the result of the measurement of projections of
two spins on axes $\av$ and $\bb$ is given by the average
$$P(\av,\bb)=\int d\lambda P(\lambda)A(\av,\lambda)B(\bb,\lambda).\eqno([3]2)$$
The question is: is it possible to find such classical
distribution $P(\lambda)$ that the classical average (\cite{bel}2)
is equal to the quantum mechanical expectation value for a
singlet, i.e. entangled, state of two particles:
$$\langle(\si_A\cdot\av)(\si_B\cdot\bb)\rangle=-\av\cdot\bb,\eqno([3]3)$$
where $\si_{A,B}$ are the spin operators (the Pauli matrices) for
particles A and B. In the attempt to answer this question, Bell
formulated his famous inequality and found that the inequality
must be violated because (see~\cite{bel}, p.19)
\bigskip

\centerline{\parbox{14cm}{\bf ``the quantum mechanical expectation
value cannot be represented, either accurately or arbitrary
closely, in the form ([3]2).''}}
\bigskip

Therefore
\bigskip

\centerline{\parbox{14cm}{\bf ``there must be a mechanism whereby
the setting of one measuring device can influence the reading of
another measurement, however remote. More over, he signal involved
must propagate instantaneously, so that such a theory could not be
Lorentz invariant.''}}
\bigskip

Notwithstanding how strange this statement looks to a physicist,
it was so attractive that it gave birth to a flood of experimental
work devoted to demonstration of Bell's inequality violation,
because this violation was identified with proof of the QM action
at a distance. We can illustrate this by citing the first
sentences from the introduction in ~\cite{hay}:
\bigskip

\centerline{\parbox{14cm}{\bf ``The concept of entanglement has
been thought to be the heart of quantum mechanics. The seminal
experiment by Aspect et al. \cite{asp1} has proved the "spooky"
nonlocal action of quantum mechanics by observing violation of
Bell inequality with entangled photon pairs.''}}
\bigskip

The nonlocality was identified with the violation of Bell's
inequalities\footnote{We speak about inequalities because there
are many modifications of the original Bell's inequality.},
because Bell proved this violation just for the correlation
(\cite{bel}3) for two remote particles in the entangled state. In
that case, if particle A passes the analyzer $\av$, the particle B
must know this fact and must be polarized along $-\av$. If the
analyzer for B is prepared along $\bb$ then the transmission
amplitude for particle B is $-\cos\theta$, where $\theta$ is the
angle between vectors $\av$ and $\bb$. However the same
correlation takes place even without action at a distance, when
the particles are initially polarized along the direction of the
vector $\av$, and their wave function is not entangled one, but a
simple product of wave functions of two particles. Moreover, the
term $\cos\theta$ appears in many cases, as shown further in the
article; and if this term is extracted from experimental data, it
is considered as the evidence of Bell's inequalities violation and
therefore as a proof of the QM non-locality. We will see below
that the violation of Bell's inequalities should not be identified
with QM action at a distance in such circumstances.

In this article I show that EPR paradox does not exist, because
particles can have position and momentum simultaneously; propose
experiments to prove local realism in QM; analyze some experiments
related to Bell's inequalities, and show that their results do not
contradict to local realism in QM.

In the next section I review the EPR paradox, demonstrate the weak
points in its formulation, and show how the paradox disappears,
thus adapting QM to local realism without any action at a
distance. In the third section I review Bohm-Aharonov
modification~\cite{ahb} of the EPR paradox, also show its weak
points, and define what exactly an experiment must measure to
prove the local realism, i.e. absence of the action at a distance.
In the fourth section I review
experiments~\cite{asp1,koco,asp2,asp3} with two cascade photons
decay of Ca atom, and demonstrate that the experimental results in
fact do not contradict the local realism. In the fifth section I
review several experiments with parametric down conversion and
show that the experimental results also do not contradict the
local realism, and that these results should be reinterpreted. In
the sixth section I show how the ``disease'' for searching quantum
miracles spreads to good experiments in neutron
physics~\cite{has1,has2}. After the conclusion, I also present the
history of submissions of this paper, along with various referees
reports.

\section{Criticism of EPR paradox}

In this section I show that EPR paradox appears only because of an incorrect definition of a value of a physical
quantity. If we change the definition, and I show that we must change it, the paradox disappears. Let's first follow
the presentation of the paradox in~\cite{epr}.

According to~\cite{epr} (p. 778) (the numeration of the equations
is the same as in the original papers. The number in front is the
reference to the papers):\bigskip

\centerline{\parbox{14cm}{\bf ``If $\psi$ is an eigenfunction of
the corresponding operator $A$, that is, if $$\psi'\equiv
A\psi=a\psi, \eqno([1]1)$$ where $a$ is the number, then the
physical quantity $A$ has with certainty the value $a$ whenever
the particle is in the state given by $\psi$.''}}
\bigskip

\noindent In particular, the momentum $p$ is defined for the wave
function represented by a plane wave
$$\psi=\exp(2\pi ip_0x/h),\eqno([1]2)$$,
since the eigenvalue of the momentum operator $\hp=(h/2\pi i)d/dx$
for this wave function is $p_0$.\bigskip

\centerline{\parbox{14cm}{\bf ``Thus, in the state given by Eq.
([1]2), the momentum has certainly the value $p_0$. It thus has
meaning to say that the momentum of the particle in the state
given by Eq. ([1]2) is real.''}}
\bigskip

In such state, however, we cannot say anything about particle's position. According to EPR~\cite{epr} we can
\bigskip

\centerline{\parbox{14cm}{\bf ``only say that the relative
probability that a measurement of the coordinate will give a
result lying between $a$ and $b$ is
$$P(a,b)=\int\limits_a^b|\psi(x)|^2dx=b-a.\eqno([1]6)$$''
}}

\subsection{The error in the EPR paper~\cite{epr}}

One can immediately see, that $P(a,b)$ in ([1]6) is not a
probability, because it is not a dimensionless quantity and it
cannot be normalized. So, {\bf the relation ([1]6) is an error}.
This error looks like a small negligence by a genius, but the
correction of this error completely resolves the paradox. I will
show that the momentum operator does not have eigenfunctions that
can be used to describe a particle state. For $P(a,b)$ to be a
valid probability, the wave function must be a normalized wave
packet,
\begin{equation}\label{2a}
\int\limits_a^b|\psi(x)|^2dx=1,
\end{equation}
and therefore
\begin{equation}\label{2b}
P(a,b)=\int\limits_a^b|\psi(x)|^2dx\neq b-a.
\end{equation}
However, wave packets cannot be eigenfunctions of the momentum
operator $\hp$. It may therefore appear that according
to~\cite{epr} the momentum cannot be a real quantity. Such
conclusion is obviously false.

It is common in QM to assume that a wave packet can be replaced
with a plane wave by limiting the consideration of the problem to
some volume $L$, and by rewriting the eigenfunction of $\hp$ in
the form $L^{-1/2}\exp(ikx)$. Then everything looks correct.
However, the choice of a finite $L$ is equivalent to either
restricting the space by two infinite walls, or to imposing
periodic boundary conditions. In former case, the wave functions
can be only of the form $\cos(kx)$ or $\sin(kx)$, and neither of
these functions are eigenfunctions of the momentum operator. In
latter case, the plane wave must be replaced by the Bloch wave
function
\begin{equation}\label{3}
\psi(x)=\phi(x)\exp(iqx),
\end{equation}
where $\phi(x)$ is a periodic function with the period $L$, and
$q$ is the Bloch wave number. This wave function is also not an
eigenfunction of the operator $\hp=-i\hbar d/dx$. Indeed, if we
expand the function $\phi(x)$ in Fourier series,
\begin{equation}\label{3a}
\phi(x)=\sum_n\phi_n\exp(2\pi inx/L),
\end{equation}
where $n$ is integer, and $\phi_n$ are Fourier coefficients, then (\ref{3}) becomes
\begin{equation}\label{3b}
\psi(x)=\sum_n\phi_n\exp(i[q+2\pi n/L]x),
\end{equation}
which shows that $\hp$ does not have a unique value. We again come to a conclusion, that according to~\cite{epr} the
momentum is not real. If, nevertheless, we follow~\cite{epr} and use a plane wave to describe a particle, we should
reject the configuration space completely. In that case the momentum completely looses its meaning.

\subsection{The resolution of EPR paradox}
To avoid the conclusions that we have reached in the previous
paragraph, we must reject the definition of ([1]1) that the value
$a$ of a physical quantity is represented by an eigenvalue of the
operator $\hA$. Instead, we should replace it with the expectation
value,
\begin{equation}\label{3c}
a=\int\psi^+\hA\psi dx.
\end{equation}
Such definition of a physical value was introduced by von Neumann~\cite{neum}. Von Neumann had shown (see adapted
version in~\cite{neum1}) that there are no dispersion free states for position and momentum operators. The important
outcome of the expectation value definition is that for any given state $\psi(x)$ of a particle we can simultaneously
find values of all the physical quantities, including momentum and coordinate, and the non-commutativity of their
operators does not matter. The commutation relations are important for calculations in QM, but they do not forbid for
particles to simultaneously have definite values of physical quantities represented by non-commuting operators. Later
we will see that the dispersion of physical values, which accompanies the definition (\ref{3c}), is not an uncertainty.
It should be interpreted differently.

To show how important this change of the value definition is for
resolving EPR paradox, let's look at the conclusion of the EPR
paper~\cite{epr} p. 780:\bigskip

\centerline{\parbox{14cm}{\bf ``Starting with the assumption that
the wave function does give a complete description of the physical
reality, we arrived at the conclusion that two physical
quantities, with noncommuting operators, can have simultaneous
reality. We are thus forced to conclude that the
quantum-mechanical description of physical reality given by wave
functions is not complete.''}}
\bigskip

However according to the new definition, the particle can have
momentum and position defined simultaneously, and the uncertainty
relations do not prevent it. As an example we can consider a
particle in the state of the Gaussian wave packet. This packet has
precisely defined velocity, which is equivalent to the classical
shift in space per unit time, and it has precisely defined
position, which can be identified with the maximum of the wave
packet. The uncertainty relation $\Delta p\Delta x\ge \hbar/4$ has
no physical meaning~\cite{conc}. It is a purely mathematical
relation that links the width of the wave packet in space to that
of its Fourier transform. The width in the Fourier space gives the
dispersion of the momentum, but this dispersion is not an
uncertainty for the packet motion. The width in the coordinate
space shows that we deal with an extended object, and the position
of an extended object is only a matter of definition. However, if
we identify the position with the center of gravity of the wave
packet, it has an absolutely certain value without any dispersion.

Since the uncertainty relation does not forbid particle to have
position and momentum simultaneously, the EPR conclusion about
incompleteness of QM becomes invalid.

\subsection{QM and local realism}
In addition to discussing the contradiction related to measurements of quantities with non-commuting operators, EPR
paper~\cite{epr} introduced entangled states and with it the action at a distance. Indeed, if we have two particles
that fly apart after interaction and don't interact any more, their wave function can be of the form
\begin{equation}\label{sup}
\Psi(x_1,x_2)=\alpha\phi(x_1)u(x_2)+\beta\psi(x_1)v(x_2),
\end{equation}
where $|\alpha|^2+|\beta|^2=1$, all the functions are normalized,
and functions $u(x_2)$, $v(x_2)$ are orthogonal. The above form of
the wave function is called ``entangled state''. The entangled
state means that, if one finds the particle 2 in the state $u(x)$,
the state of the particle 1 immediately becomes $\phi(x)$. How
does the particle 1 knows which term in the superposition
(\ref{sup}) to choose? Does it mean that there is an instantaneous
transfer of information from particle 2 to particle 1? If the
answer is ``yes'', then it means that there is an instant action
at a distance and that QM is a nonlocal theory.

I reject such possibility and view the entangled state as a set of possible product states in a realization of a single
event in an ensemble of many events. That is, in a single event the state of two particles is either
\begin{equation}\label{sup1}
\Psi(x_1,x_2)=\phi(x_1)u(x_2),
\end{equation}
or
\begin{equation}\label{sup2}
\Psi(x_1,x_2)=\psi(x_1)v(x_2),
\end{equation}
and both states appear with frequencies (probabilities) $\alpha$ and $\beta$. The state that appears in every given
realization is determined by a hidden parameter of QM (do not mistake it with a classical hidden parameter). This
hidden parameter does not allow to predict all physical quantities without quantum probabilities.

\section{Bohm-Aharonov formulation of the EPR paradox}

There are no feasible experiments that can be used to check the original EPR paradox. A modified version of the paradox
is presented in~\cite{ahb}. In this version of the paradox we can clearly demonstrate that the non-commutativity does
not matter.

Instead of the EPR scattering experiment, the paper~\cite{ahb} considers the decay of a scalar molecule into two atoms
with spins 1/2. Because of Fermi statistics the spin wave function of the two atoms is represented by an entangled
state
$$\psi=\fr1{\sqrt 2}[\psi_+(1)\psi_-(2)-\psi_-(1)\psi_+(2)],\eqno([2]1)$$
where $\psi_\pm(1)$ refers to the states of the atom A where its spin projections on an arbitrary selected axis is
equal to $\pm1/2$, and $\psi_\pm(2)$ refers to states of the atom B with its spin projections $\pm1/2$ on the same
axis.

After the atoms separate by a sufficient distance,
\bigskip

\centerline{\parbox{14cm}{\bf ``...so that they cease to interact, any desired component of the spin of the first
particle (A) is measured. Then, because the total spin is still zero, it can be concluded that the same component of
the spin of the other particle (B) is opposite to that of A. ''}}
\bigskip

The paper then continues ``If this were a classical system there would be no difficulties, in interpreting the above
result, because all components of the spin of each particle are well defined at each instant of time.'' However in
quantum theory there is a problem:
\bigskip

\centerline{\parbox{14cm}{\bf ``In quantum theory, a difficulty
arises, in the interpretation of the above experiment, because
only one component of the spin of each particle can have a
definite value at a given time. Thus, if the $x$ component is
definite, then $y$ and $z$ components are indeterminate and we may
regard them more or less as in a kind of random fluctuation.''}}
\bigskip

It means that there must be action at a distance, otherwise the second particle would not know which one of the spin
components ceases to fluctuate when we measure the spin component of the first particle. I will now show that there is
no action at a distance because all components of the spin have certain values and therefore no spin fluctuations are
present. For a spin state $|\psi\rangle$ the ``spin arrow'' $\sbb$ is defined as
\begin{equation}\label{sar1}
\sbb=\langle\psi|\si|\psi\rangle,
\end{equation}
where $\si=(\sigma_x,\sigma_y,\sigma_z)$, and $\sigma_{x,y,z}$ are
well known Pauli matrices. Every atom has some polarization.
Suppose a particle is in a state described by a
spinor\footnote{This representation is valid for an arbitrary unit
vector $\OO$ except for $\OO=(0,0,-1)$, but a single point doesn't
matter.}
\begin{equation}\label{I1}
|\psi_{\OO}\rangle=\fr{1+\si\OO}{\sqrt{2(1+\Omega_z)}}{1\choose0},
\end{equation}
where $\OO$ is a unit vector $(\sin\theta\cos\varphi,\sin\theta\sin\varphi,\cos\theta)$, $\theta,\varphi$ are polar
angles, and the quantization axis is chosen along $z$. The states
\begin{equation}\label{I2}
|\psi_{\pm\OO}\rangle=\fr{1\pm\si\OO}{\sqrt{2(1\pm\Omega_z)}}{1\choose0},
\end{equation}
are the normalized eigenstates of the matrix $\si\OO$ with eigenvalues $\pm1$, i.e.
$\si\OO|\psi_{\pm\OO}\rangle=\pm|\psi_{\pm\OO}\rangle$. Spin arrow $\av$ in these states is defined as a unit vector
\begin{equation}\label{I3}
\sbb=\langle|\psi_{\Omega}|\si|\psi_{\Omega}\rangle=\OO.
\end{equation}
This vector is equal to the unit vector $\OO$ along which the
particle is polarized. Thus it is absolutely true that if we could
measure a component $\sbb_x=\OO_x$ of the particle A, we would
immediately predict with certainty that the projection $\sbb_x$ of
the spin-arrow of the particle B is precisely $-\OO_x$, and all
the other components have precisely defined values. The components
do not fluctuate, even though we do not measure them. Spinors and
spinor states are the mathematical objects, which are used as a
building blocks for constructing a classical quantity --- spin
arrow.

In reality however we cannot measure the value of a spin arrow component. The only thing we can do is to count
particles, and use our measuring devices as filters. A measurement means that we have an analyzer, which transmits
particles with, say, $x$ components only. In addition, filtering modifies the state of polarization of the particle. If
the particle is in the state (\ref{I1}), then after the filter, which is oriented along $\sbb_x$-component, the state
becomes
\begin{equation}\label{J1}
|\psi_f\rangle=\fr{1+\sigma_x}{2}|\psi_\Omega\rangle=
\fr{1+\Omega_z+\Omega_x+i\Omega_y}{2\sqrt{1+\Omega_z}}|\psi_x\rangle\equiv
f_x|\psi_x\rangle,
\end{equation}
where
\begin{equation}\label{J1J1}
f_x= \fr{1+\Omega_z+\Omega_x+i\Omega_y}{2\sqrt{1+\Omega_z}}.
\end{equation}
The probability of the filter transmission is equal to
\begin{equation}\label{J2}
w(\OO\to x)=|f_x|^2=\fr{1+\Omega_x}{2},
\end{equation}
which is very evident. And of course while particle 1 after the filter becomes in the state $\psi_x$, the second one
remains in the state $\psi_{-\OO}$, and this fact does not contradict to anything.

We can easily construct the wave function of two spinor particles inside the primary scalar one, as an isotropic
superposition of particles with the opposite spin arrows. Indeed, the state ([2]1) can be represented as
\begin{equation}\label{J3}
\psi=\fr1{\sqrt2}\int\limits_{2\pi}\fr{d\Omega}{4}\Bigg[
\fr{(1+\si\OO)_\xi}{\sqrt{2(1+\Omega_z)}}
\fr{(1-\si\OO)_\chi}{\sqrt{2(1-\Omega_z)}}-\fr{(1-\si\OO)_\xi}{\sqrt{2(1-\Omega_z)}}
\fr{(1+\si\OO)_\chi}{\sqrt{2(1+\Omega_z)}}\Bigg]|\xi_+\chi_+\rangle,
\end{equation}
where $\xi,\chi$ denote particles that after the decay fly to the right and left respectively. Indices $\xi,\chi$ of
the numerators in the fractions show that the operators should act upon the spinor states $|\xi_+\rangle$,
$|\chi_+\rangle$, which are identical and equal to
$$|\xi_+\rangle=|\chi_+\rangle={1\choose0}.$$
The integration in (\ref{J3}) is done over the solid angle $2\pi$, i.e. over the right hemisphere. This limitation is
necessary to preserve the antisymmetry of the spinor wave function and to preserve the total spin of the scalar
particle equal to zero instead of a superposition of states with spins 0 and 1.

The expression (\ref{J3}) can be easily verified. It can be simplified and reduced to
\begin{equation}\label{J4}
\psi=\fr1{\sqrt2}\int\limits_0^{\pi}d\theta\int\limits_{-\pi/2}^{\pi/2}\fr{d\varphi}{4}
[(\si\OO)_\xi-(\si\OO)_\chi]|\xi_+\chi_+\rangle=\fr1{\sqrt2}\bigg[|\xi_-\chi_+\rangle-
|\xi_+\chi_-\rangle\bigg],
\end{equation}
which is identical to ([2]1).

The decay is a transition from the entangled state with the total
spin zero to a product state with the zero total spin projection
onto any direction. The direction of the flight breaks the
isotropy of the space and opens the door for the total angular
momentum non-conservation. After the decay, the two particles fly
apart with opposite polarizations parallel to some unknown
direction\footnote{The probability amplitudes have different signs
with respect to the interchange of the particles' flight
directions. However the sign does not affect the probability
itself.} $\pm\OO$. If we were so lucky as to guess direction $\OO$
of the spin arrow of the first particle $|\xi\rangle$ going to the
right, and orient our measuring apparatus precisely along it, we
would be able with absolute certainty predict the direction and
therefore all its components of the spin arrow of the second
particle $|\chi\rangle$ going to the left. However usually we are
not so lucky, and we cannot see the spin arrow of a given
particle. We can only use a filter, for example a magnetic field,
which is directed, say, along vector $\bb$, and transmits
particles polarized along $\bb$. The transmitted particles are
then registered by a detector. Particles polarized in opposite
directions are reflected by the analyzer and are not registered by
the detector. Particles polarized in other directions are
transmitted by this analyzer only with some probability. If the
atoms are polarized along $\OO$, the probability is equal to
\begin{equation}\label{J6}
w=\langle\psi_{\Omega}|\fr{1+\si\bb}{2}|\psi_{\Omega}\rangle=
\fr{1+(\bb\OO)}{2}.
\end{equation}

After the transmission, particle A becomes in the state
$|\psi_{\bb}\rangle$, but it does not mean that the particle B,
which is going to the left, is in the state $|\psi_{-\bb}\rangle$.
It remains in the state $|\psi_{-\Omega}\rangle$, and one can
measure (filter) it along any direction. We can thus find the
transmission probability for particle B through any filter (if we
ever could measure the probability of a single event).

\subsection{How to prove the local realism experimentally}

To prove the local realism experimentally we must measure the ratio between the coincidence count rates for parallel
and antiparallel analyzers.

Let's consider a scalar molecule in the state ([2]1), decaying into two spin 1/2 atoms. If we imagine two parallel
analyzers inside the molecule before the decay, we can prove that their coincidence count rate is zero. Indeed, suppose
the analyzers are oriented along the direction $\OO$. Then the mutual transmission probability is
$$w=\Big[\langle\xi_{-}\chi_{+}|-\langle\xi_{+}\chi_{-}|\Big]\fr{(1+\si\OO)_\xi}{2}
\fr{(1+\si\OO)_\chi}{2}\Big[|\xi_{-}\chi_{+}\rangle-|\xi_{+}\chi_{-}\rangle\Big]=$$
$$\Bigg[\langle\xi_{-}\chi_{+}|\fr{(1+\sigma_z\Omega_z)_\xi}{2}
\fr{(1+\sigma_z\Omega_z)_\chi}{2}|\xi_{-}\chi_{+}\rangle+\langle\xi_{+}\chi_{-}|
\fr{(1+\sigma_z\Omega_z)_\xi}{2}
\fr{(1+\sigma_z\Omega_z)_\chi}{2}|\xi_{+}\chi_{-}\rangle\Bigg]-$$
$$-\Bigg[\langle\xi_{-}\chi_{+}|\fr{(\sigma_x\Omega_x+\sigma_y\Omega_y)_\xi}{2}
\fr{(\sigma_x\Omega_x+\sigma_y\Omega_y)_\chi}{2}|\xi_{+}\chi_{-}\rangle+$$
$$+\langle\xi_{+}\chi_{-}|\fr{(\sigma_x\Omega_x+\sigma_y\Omega_y)_\xi}{2}
\fr{(\sigma_x\Omega_x+\sigma_y\Omega_y)_\chi}{2}|\xi_{-}\chi_{+}\rangle\Bigg]=$$
\begin{equation}\label{J7}
=\fr{(1-\Omega_z^2)-|\Omega_x+i\Omega_y|^2}{2}=\fr{1-\Omega^2}{2}=0.
\end{equation}
The details in the above calculations show that zero is obtained
only because of the interference, which is represented by the
contribution of the cross terms.

Therefore, in a scalar molecule represented by an entangled state
of two spinor atoms, the probability to find both atoms with the
same direction of the spin is zero. However what will happen after
decay? Before the decay there are no left and right sides. After
the decay the state of the two atoms changes. The atoms acquire a
new parameter --- direction of their flights.  So, if we position
parallel analyzers far apart on the opposite sides of the
molecule, will we see some coincidence in count rate of atoms? The
answer is ``yes''.

Indeed, after the decay the two particles go apart with the
arbitrary direction of the spin arrow. This direction is a hidden
parameter. It is a unit vector homogeneously distributed over half
of the sphere. If we were able to measure it we would have a
classical picture.

The wave function can be represented by (\ref{J3}), and the
transmission probability through two analyzers oriented along
vectors $\av$ and $\bb$ is
\begin{equation}\label{J8}
w=\int\fr{d\Omega}{4\pi}\langle\xi_+\chi_+|\fr{(1-\si\OO)_\xi}{\sqrt{2(1-\Omega_z)}}
\fr{(1+\si\OO)_\chi}{\sqrt{2(1+\Omega_z)}}\fr{(1+\si\av)_\xi}{2}\fr{(1+\si\bb)_\chi}{2}
\fr{(1-\si\OO)_\xi}{\sqrt{2(1-\Omega_z)}}
\fr{(1+\si\OO)_\chi}{\sqrt{2(1+\Omega_z)}}|\xi_+\chi_+\rangle.
\end{equation}
Since the particles are far apart there are no interference cross
terms, and the antisymmetry plays no role in calculating the
probabilities.

With the help of (\ref{J6}) we immediately obtain
\begin{equation}\label{J9}
w(\av,\bb)=\int\fr{d\Omega}{4\pi}\fr{[1-(\OO\av)]}{2}\fr{[1+(\OO\bb)]}{2}.
\end{equation}
This relation is just what can be expected in classics. It has the
well known form for a hidden parameter $\lambda$:
\begin{equation}\label{J10}
w(\av,\bb)=\int\rho(\lambda)w(\lambda,\av)w(\lambda,\bb),
\end{equation}
where $\rho(\lambda)$ is the probability distribution for the
hidden parameter, and $w(\lambda,\bb)$, $w(\lambda,\av)$ are
probabilities to count the particle when the analyzers are
oriented along $\av$ and $\bb$.

Simple calculations give
\begin{equation}\label{JJ9}
w(\av,\bb)=\fr14\left(1-\fr{\av\bb}{3}\right).
\end{equation}
Therefore
\begin{equation}\label{J11}
w(\av,\av)=1/6,\qquad w(\av,-\av)=1/3,
\end{equation}
and, instead of measuring the violation of Bell's inequalities, one should check if
\begin{equation}\label{J12}
\fr{w(\av,\av)}{w(\av,-\av)}=\fr12,
\end{equation}
to prove or reject local realism

We see that the coincidence count rate for parallel analyzers is
only two times smaller than for antiparallel ones (according to
EPR, the former is thought to be zero). It is easy to understand.
The atoms after the decay go apart with antiparallel
polarizations. But the axis of these polarizations is random. If
both analyzers filter the same direction, the probability for both
atoms to be registered is larger than zero, since their
polarizations are in general at an angle with respect to the
analyzers.

\subsection{Experiments with photons}

Most of the initial experiments were however done with photons,
not atoms. Photons are also considered in~\cite{ahb}, but mainly
with respect to the annihilation of electron-positron pairs. At
the time, the experiments with cascade decay of an excited atom
into two photons were most practical.

\section{Experiment with cascade photons}
\subsection{The first experiment with photons}

The first such experiment is presented in~\cite{koco}. The authors
measured photons from the cascade 6$^1S_0\to4^1P_1\to4^1S_0$ in
calcium. To understand what correlations between the photon
polarizations can be expected, the authors write:
\bigskip

\centerline{\parbox{14cm}{\bf ``Since the initial and final atomic
states have zero total angular momentum and the same parity, the
correlation is expected to be of the form
$(\bm{\epsilon_1\cdot\epsilon_2})^2$. ''}}
\bigskip

It means that the wave function of the photons inside the atom can be represented as
\begin{equation}\label{n1}
|\psi\rangle=\fr1{\sqrt2}(C^xC^x+C^yC^y)|\psi_0\rangle,
\end{equation}
where $|\psi_0\rangle$ is the vacuum state and $C^{x,y}$ are the
creation operators for photons with polarizations along $x$ or $y$
directions. Also, we can rewrite the wave function (\ref{n1}) of
the two photons in the form similar to (\ref{J4}),
\begin{equation}\label{n2}
|\psi\rangle=\fr1{\sqrt2}[|x,x\rangle+|y,y\rangle].
\end{equation}
However since the total electric field of the two photons should
be zero, their function is better to write as
\begin{equation}\label{n2a}
|\psi\rangle=\fr1{\sqrt2}[|x,-x\rangle+|y,-y\rangle],
\end{equation}
where the first letter is related to the photon going to the
right, and the second one --- to the left. It is clear that we can
also write it as a uniform distribution
\begin{equation}\label{n2b}
|\psi\rangle=\fr1{\sqrt2}\int\fr{d\phi}{2\pi}|\ks(\phi),\cb(\pi+\phi)\rangle,
\end{equation}
where $\ks$ and $\cb$ denote particles going to the right and left
respectively, similar to (\ref{J3}). Before the decay, the
projection of this wave function on two filters oriented along
$\av$ and $\bb$ is
\begin{equation}\label{n2c}
\langle\av_\xi,\bb_\chi||\psi\rangle=\fr1{\sqrt2}\int\fr{d\phi}{2\pi}\langle\av|
|\ks(\phi)\rangle\langle\bb||\cb(\pi+\phi)\rangle=
\fr1{\sqrt2}\int\fr{d\phi}{2\pi}\cos(\phi)\cos(\phi+\pi-\varphi),
\end{equation}
where $\varphi$ is the angle between $\av$ and $\bb$. After integration we get
\begin{equation}\label{n2d}
\langle\av_\xi,\bb_\chi||\psi\rangle=-\fr1{\sqrt2}\cos(\varphi),
\end{equation}
which shows that the projection is zero for $\varphi=\pi/2$. Probability of such projections is
\begin{equation}\label{nn2d}
|\langle\av_\xi,\bb_\chi||\psi\rangle|^2=\fr{1+\cos(2\varphi)}{4}.
\end{equation}
Now we will make similar calculations for coincidence counts after
the decay.

\subsubsection{Derivation of the ratio}

After decay the transmission probability of photons through two
filters located on left and right and oriented along $\av$ and
$\bb$ respectively is
\begin{equation}\label{n2e}
P(\av,\bb)=\int\fr{d\phi}{2\pi}|\langle\av|
|\xi(\phi)\rangle|^2|\langle\bb||\chi(\pi+\phi)\rangle|^2=
\int\fr{d\phi}{2\pi}\cos^2(\phi)\cos^2(\phi-\varphi)=\fr14\bigg[1+
\fr{\cos(2\varphi)}{2}\bigg].
\end{equation}
We immediately see that for $\varphi=0$ or $\pi$ the probability
is 3/8, and for $\varphi=\pi/2$ it is 1/8. So the predicted ratio
is
\begin{equation}\label{prr}
\fr{w(\pi/2)}{w(0)}=\fr13.
\end{equation}
It is very important to notice that the probability of the
coincidence counts with orthogonal filters is not zero. It is only
three times less compared to the coincidences with parallel
filters. It is also important to note that (\ref{n2e}) can be also
represented as
\begin{equation}\label{n2en}
P(\av,\bb)=\fr12\left(\fr14+\fr{1+\cos(2\varphi)}{4}\right),
\end{equation}
so it differs from the result (\ref{nn2d}) only by a constant!
\begin{figure}[!h]
{\par\centering{\includegraphics[width=\textwidth]{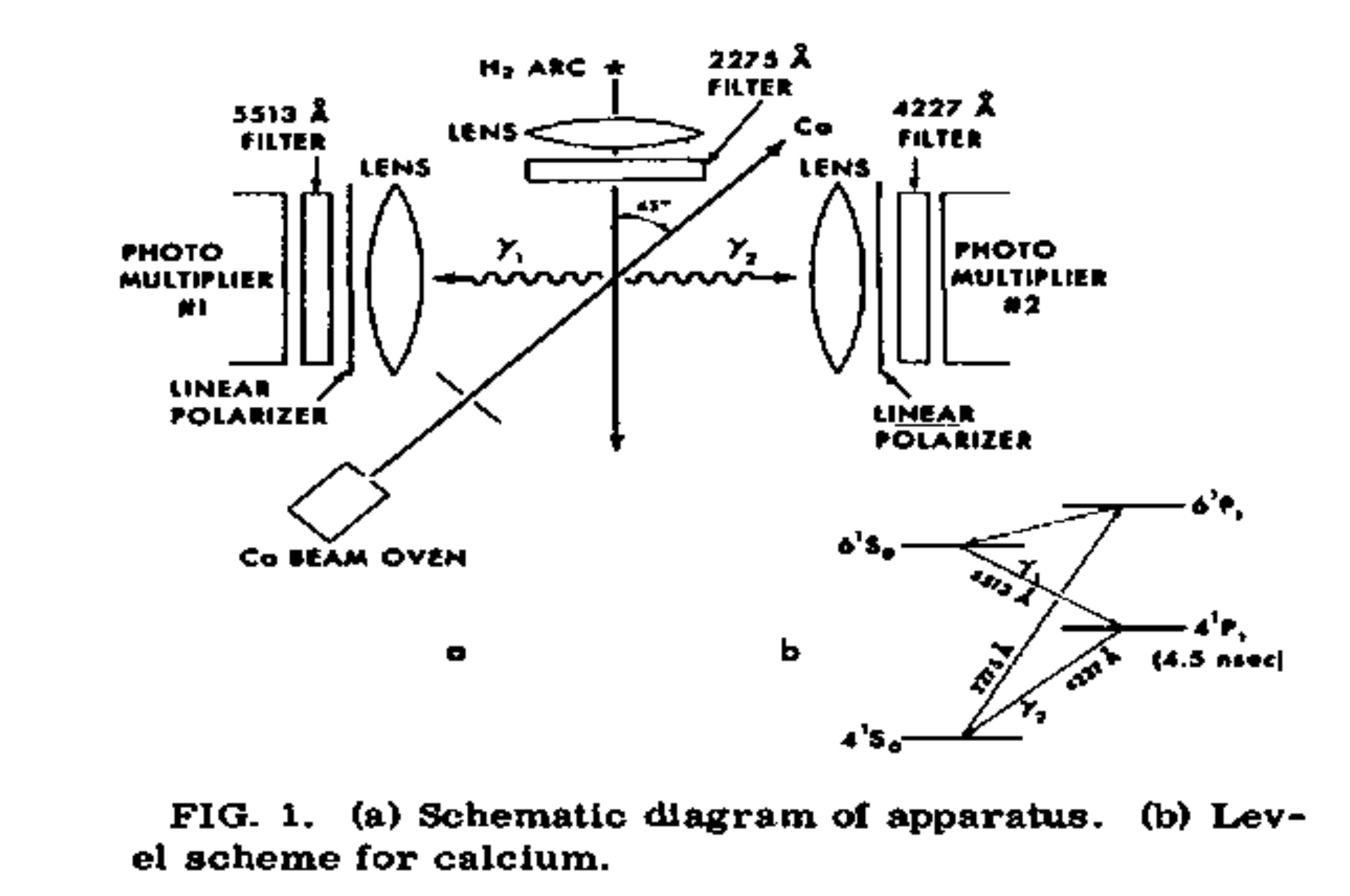}}
\par} \caption{Scheme of the first experiment with cascade photons~\cite{koco}. \label{koch5}}
\end{figure}

\begin{figure}[!h]
{\par\centering\resizebox*{12cm}{!}{\includegraphics{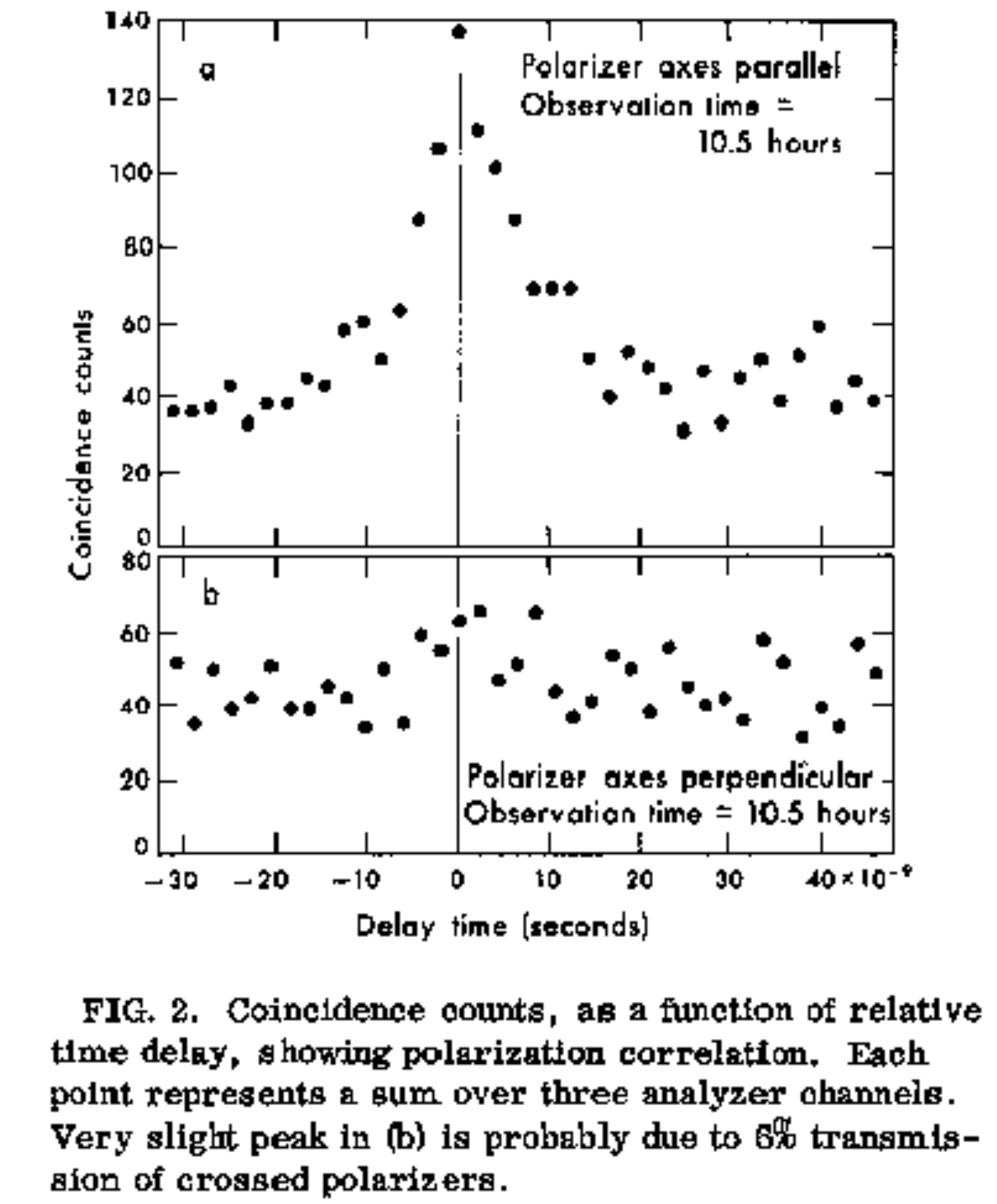}}
\par} \caption{Coincidence counts at different detector delays
obtained in the experiment by~\cite{koco}. \label{1st1}}
\end{figure}
\subsubsection{Experimental results}
Let's review the experiment by~\cite{koco} (Fig. \ref{koch5}). The
results of this experiment are demonstrated in fig. \ref{1st1},
where number of counts after parallel and perpendicular analyzers
are shown in dependence of delay in coincidence times. We clearly
see that the background can be accepted at the level 40 counts,
therefore at maximum the coincidence counts with parallel filters
is equal to $w(0)\approx100$, and the coincidence counts with
orthogonal filters is $w(\pi/2)\approx30$. Thus the measurement
precision in this experiment allows for a good agreement with
$w(\pi/2)/w(0)\approx1/3$.

However the authors, although absolutely honest, are very
reluctant to accept the local realism interpretation. Their
conclusion is:
\bigskip

\centerline{\parbox{14cm}{\bf ``The results of a 21-h run, shown
in Fig. \ref{1st1}. indicate clearly the difference between the
coincidence rates for parallel and perpendicular orientations.
They are consistent with a correlation of the form
$(\bm{\epsilon_1\cdot\epsilon_2})^2$. ''}}
\bigskip

\noindent And they attribute the discrepancy at 30\% level to 6\%
transmission of crossed filters, as is documented in the caption
of fig. \ref{1st1}. I do not accept such explanation as a proof
for action at a distance, while the majority does.

\subsection{The experiments by Aspect a.o.}

\begin{figure}[!h]
{\par\centering\resizebox*{12cm}{!}{\includegraphics{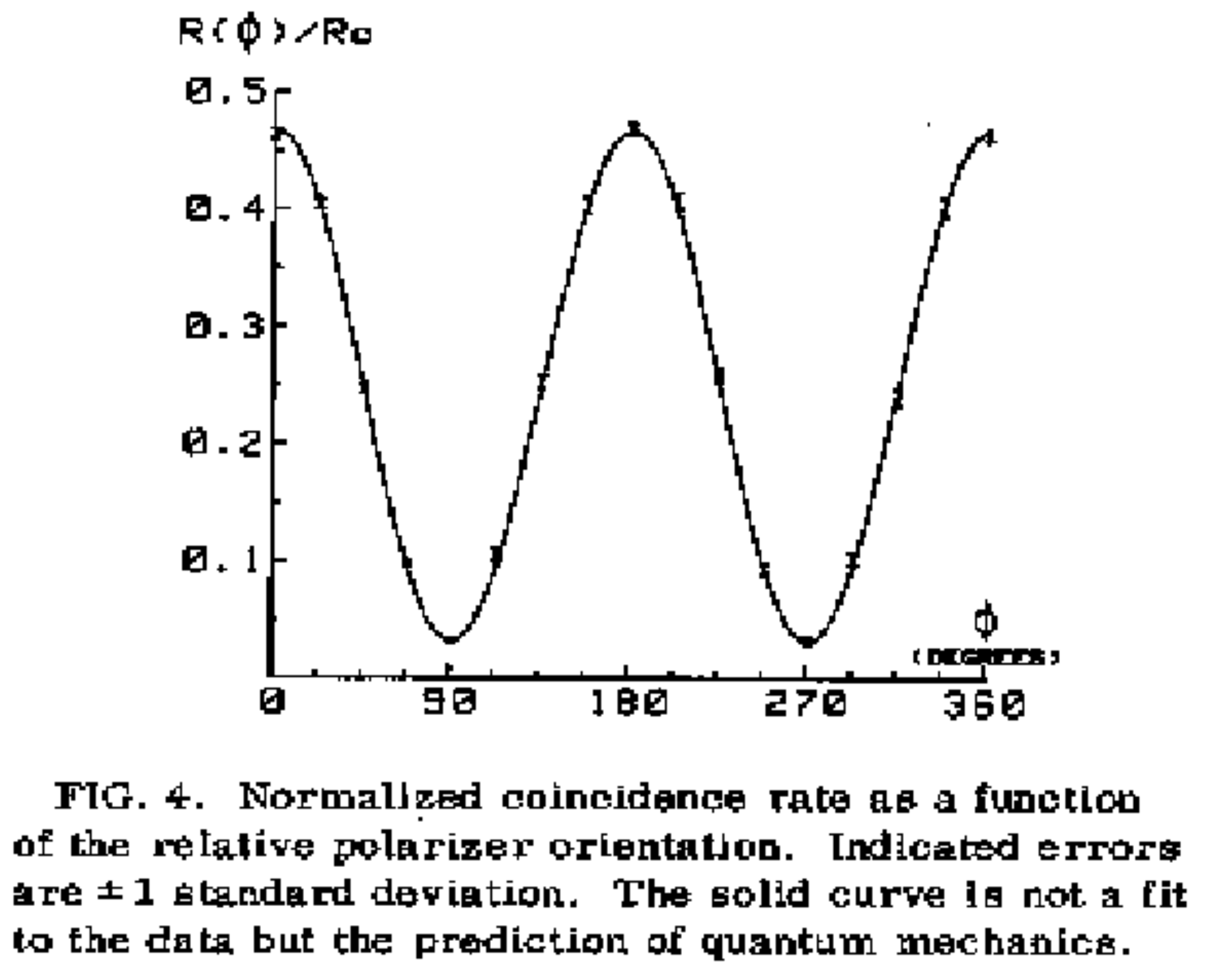}}
\par} \caption{Results of the experiment~\cite{asp1}. \label{asp7}}
\end{figure}

The experiments by Aspect a.o., in particular the first
one~\cite{asp1}, are commonly accepted as an experimental proof of
QM action at a distance. The experiment~\cite{asp1} is very
similar to the one described in~\cite{koco} but instead of the 6
$^1S_0$ state they excited the $4p^2$ $^1S_0$ state, and they
measured the coincidence rates at different angles between the two
analyzers. Let's look at the experimental results, shown in Fig.
\ref{asp7}.

This result can be described as (\ref{nn2d}) and not as
(\ref{n2e}), however the difference is only in a constant term
1/8, and in normalization. The normalization is not important as
it is more or less arbitrary, but the constant term is crucial.
Let's look how the authors subtracted the constant term:
\bigskip

\centerline{\parbox{14cm}{\bf ``Typical coincidence rates without
polarizers are 240 coincidences per second in the null delay
channel and 90 accidental coincidences per second; for a 100-s
counting period we thus obtain 150 true coincidences per second
with a standard deviation less than 3 coincidences per second.''}}
\bigskip

Suppose that not all 90 coincidences are accidental but only 50,
while the remaining 40 are not accidental. We then immediately
find that at the angle $\varphi=\pi/2$ between the polarizers
there are 40 coincidence counts, and 115 coincidence counts at
$\varphi=0$ (ratio $\approx1/3$). In this case there is no
contradiction to quantum mechanics, and there is no proof of the
spooky action at a distance. Everything is in accordance with our
calculations and local realism.

In another experiment by the same authors~\cite{asp2}, where the
photons are registered by two detectors on each side in order not
to lose particles of different polarizations, the background
coincidences are also subtracted as accident ones. The decision to
consider the counts as accidental, and the number of such counts
are cited in the following paragraph,
\bigskip

\centerline{\parbox{14cm}{\bf ``Each coincidence window, about 20
ns wide, has been accurately measured. Since they are large
compared to the lifetime of the intermediate state of the cascade
(5 ns) all true coincidences are registered. We infer the
accidental coincidence rates from the corresponding single rates,
knowing the width of the windows. This method is valid with our
very stable source, and it has been checked by comparing it with
the methods of Ref. 5, using delayed coincidence channels and/or a
time-to-amplitude converter. By subtracting of these accidental
rates (about 10 s$^{-1}$) from the total rates, we obtain the true
coincidence rates $R_{\pm,\pm}(\av,\bb)$ (actual values are in the
range 0--40 s$^{-1}$, depending on the orientations).''}}
\bigskip

From this citation we see again that the subtracted number of
coincidences is in the order of value that can also be expected
for orthogonal directions of analyzers. Therefore, both
experiments~\cite{asp1,asp2} must be re-evaluated.

The last experiment~\cite{asp3} is of the ``delayed choice'' type.
I think that the delayed choice experiments do not check anything,
since the spooky action at a distance is absent. In conclusion we
can say that the results of all cascade experiments are not
indisputable enough to reject the local realism.

\section{Bell's inequalities in down-conversion experiments}

In the last decade many of experiments that were aimed at proving
the violation of Bell's inequalities were done with spontaneous
parametric down-conversion (SPDC) photons. In these experiments,
initially polarized photon with frequency $\omega$ splits into two
polarized photons with frequencies $\omega/2$. The generated
photons are thought to be in entangled state, which can be
modified with the help of optical devices, and can be used to show
the violation of Bell's inequalities. In the next part of this
article we will review three types of such experiments, which use
three different approaches to prove the violation --- a) photon
polarization measurements, b) travel path interference and c)
counts statistics. I think that these experiments are well
designed, highly resourceful, and can be used to learn a lot about
physics of the photon down-conversion, but instead they are
narrowly aimed at what appears to be a useless quest
--- demonstration of the Bell's inequalities violation.

\subsection{Polarization experiments~\cite{kwi99}.}

\begin{figure}[!h]
{\par\centering\resizebox*{12cm}{!}{\includegraphics{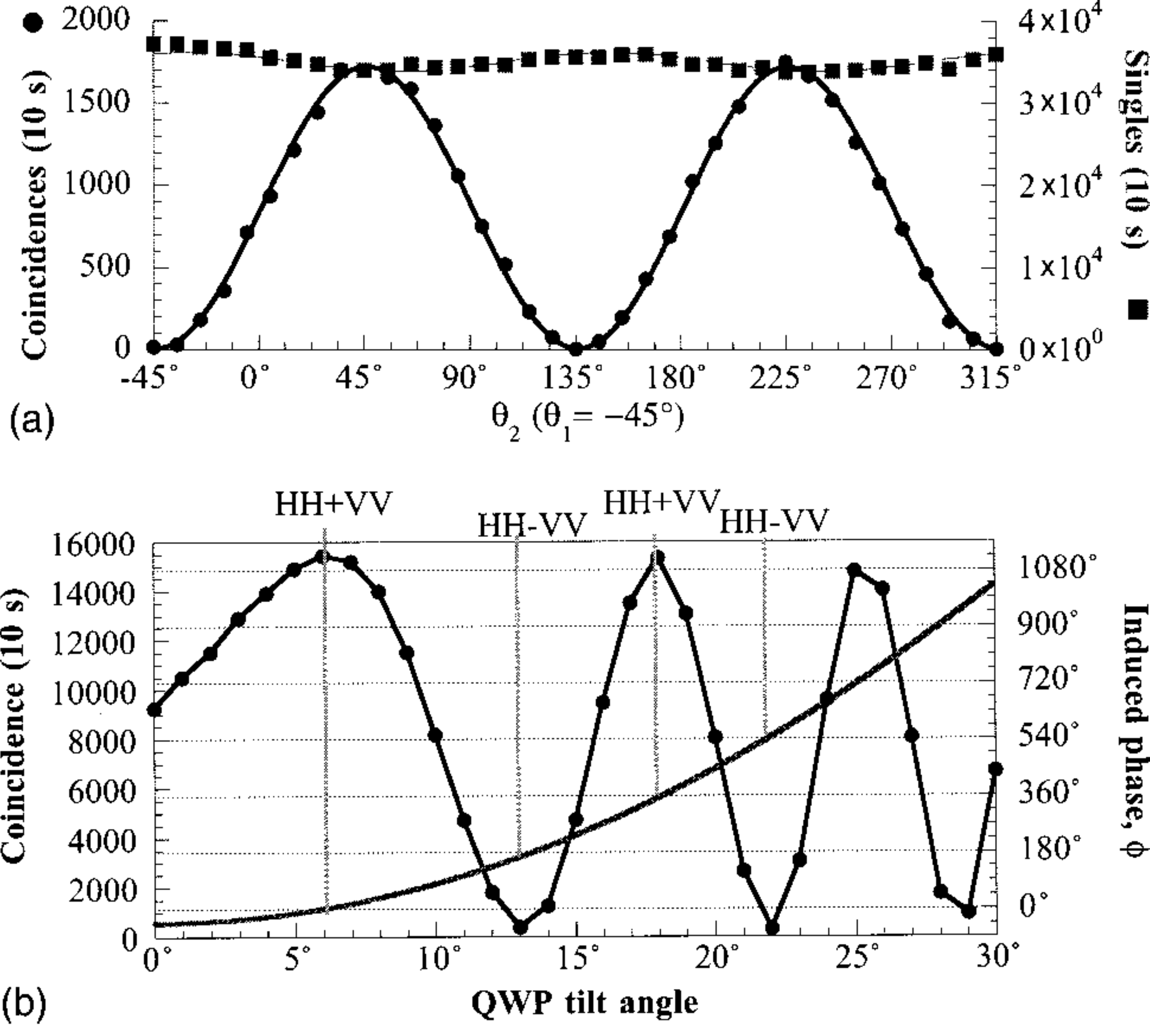}}
\par} \caption{\label{kwi3} a) Measurements of the polarization entanglement. The
polarization analysis of photon 2 was varied, while that of photon
1 was at -45-. The rate at detector 2 ~squares, right axis! is
essentially constant; i.e., the photons are individually nearly
unpolarized, while the coincidence rate ~circles, left axis!
displays the expected quantum-mechanical correlations. The solid
curve is a best fit, with visibility $V=99.66\pm0.3$\% ~b)
Coincidences as the relative phase $\varphi$ was varied by tilting
the wave plate just before the crystal; both photons were analyzed
at 45-. The solid curve is the calculated phase shift for our
2-mm-thick zero-order quartz quarter-wave plate, adjusted for the
residual phase shift from the BBO crystals themselves.}
\end{figure}

The schematic of the experiment is shown in Fig.~\ref{fkwi}. The
details of this experiment and the types of the generated
entangled states are not included here, because this information
is not important in the scope of this work. Instead, we focus on
the result presented in Fig.~\ref{kwi3}. The authors say:

\begin{figure}[!t]
{\par\centering\resizebox*{12cm}{!}{\includegraphics{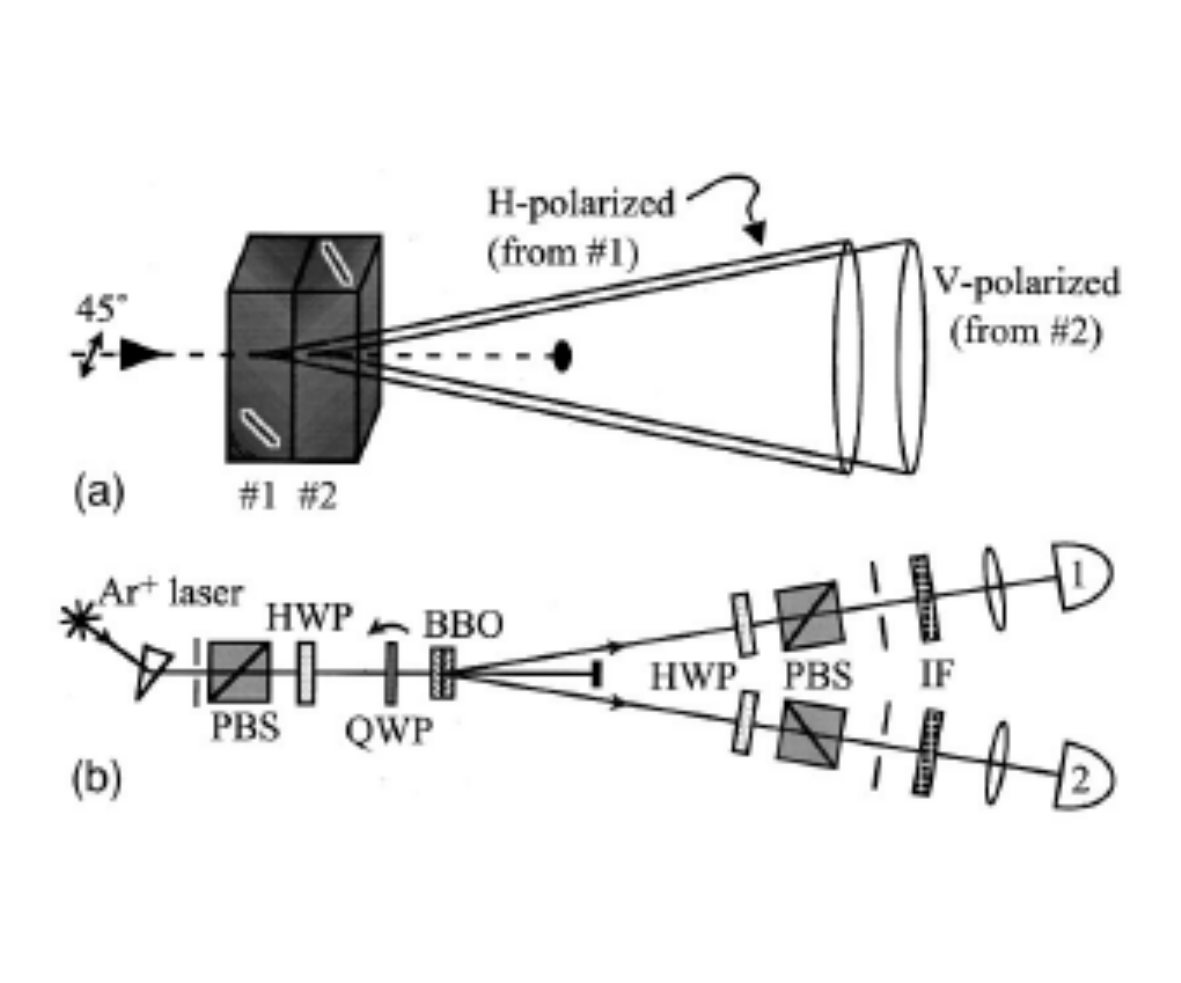}}
\par}  \caption{\label{fkwi} Generation of the entangled state by the down-conversion of laser photons
in two adjacent nonlinear crystals with perpendicular crystalline
axes~\cite{kwi99}.}
\end{figure}
\medskip

\centerline{\parbox{14cm}{\bf ``Figure \ref{kwi3} shows data
demonstrating the extremely high degree of polarization
entanglement achievable with our source. The state was set to
$HH-VV$ the polarization analyzer in path 1 was set to
$-45^\circ$, and the other was varied by rotating the HWP in path
2.''}}
\medskip
\medskip

The result in Fig. \ref{kwi3} (upper part) can be easily
reproduced without taking into account the entanglements and the
action at a distance. Instead, the same data can be acquired if we
assume that the photon 1 is polarized at the angle -45$^\circ$,
and the photon 2 is at the angle +45$^\circ$. Furthermore, if both
polarizations rotate when the angle of the quarter-wave plate is
changed, the lower part of Fig. \ref{kwi3} is also well
reproduced. In order to reject such description, the authors
should have measured the photons polarizations in the two
channels. Such test was not done. The authors limit themselves to
noting that the count rate at a single detector does not depend on
the orientation of the analyzer (upper part of fig. \ref{kwi3}).
Notice, however, that the count rate of single counts is at least
20 times higher than the rate of the coincidence counts. No
explanation is given to such high level of singles.
(See~\cite{kwi99}.)

In order to reject the above suggestions, the polarization
measurements should be done also at coincidence counting. This was
not done, therefore the experimental results in~\cite{kwi99} are
incomplete and cannot be interpreted in favor of the action at a
distance. The same conclusion can be drawn about~\cite{kwi95}
and~\cite{zei}.

\subsection{Path interference experiments~\cite{kwi93}}

\begin{figure}[!h]
{\par\centering\resizebox*{12cm}{!}{\includegraphics{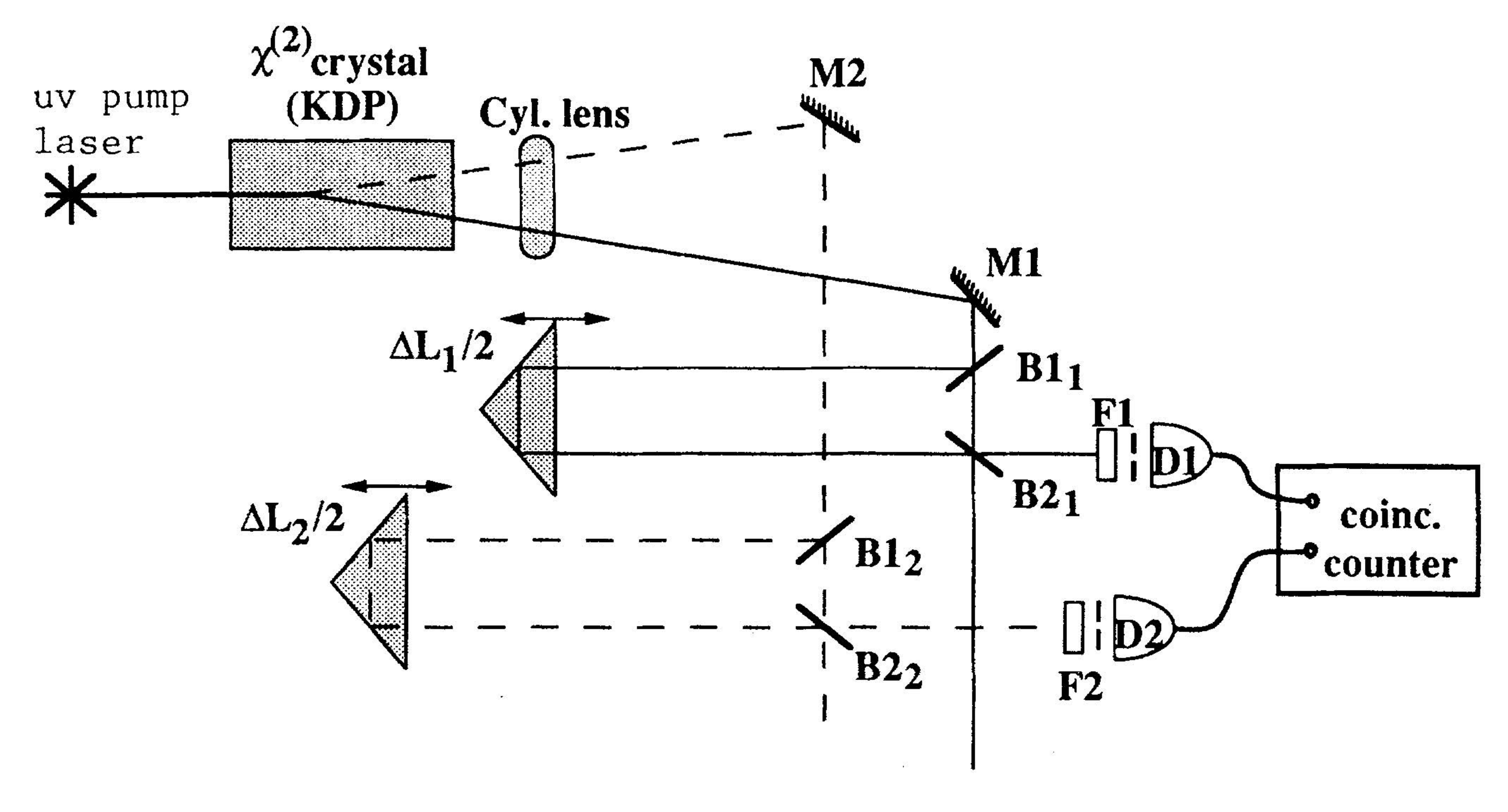}}
\par} \caption{\label{kwi93} Schematic of the experiment used in~\cite{kwi93}.}
\end{figure}

The schematic of the path interference experiment~\cite{kwi93} is shown in Fig. \ref{kwi93}

The photons can travel to each detector along two possible paths
--- a short and a long ones, and the interference of these two
paths is measured. It is assumed that the photon travelling to the
second detector receives the information about the path chosen by
the photon travelling to the first detector, and chooses the same
long or short path. If the short path, identical for both
detectors, is fixed, and the long one is identically varied, the
coincidence count rate periodically changes. This periodic
behavior is then used to prove the violation of Bell's
inequalities. However this interference has a very simple
alternative explanation, which does not require the action at a
distance. When we measure the coincidence counts at time $t_0$, we
count photons created in the crystal at times $t_0-t_s$ and
$t_0-t_l$, where $t_{s,l}$ are the times of flight for photons
from the creation point to the detectors along the long and short
paths respectively. Since the photons are created by a coherent
wave, then the decrease of the count rate at $\omega(t_l-t_s)=\pi$
is related to the coherent suppression of the creation of such
photons, and not with the interference between the long and short
paths. In order to investigate and prove this effect, it is
necessary to fix the path difference at one detector and to vary
one of the paths to the second one. In this case we could study
the interference of the photon down-conversion at three points.

We can conclude that this experiment has nothing to do with the
EPR paradox and the action at a distance. The same can be said
about experiment~\cite{pit}, where the geometrical change of paths
was replaced by time delays.

\subsection{Experiments with counting statistics of photons~\cite{ou}}

\begin{figure}[!h]
{\par\centering{\includegraphics[width=\textwidth]{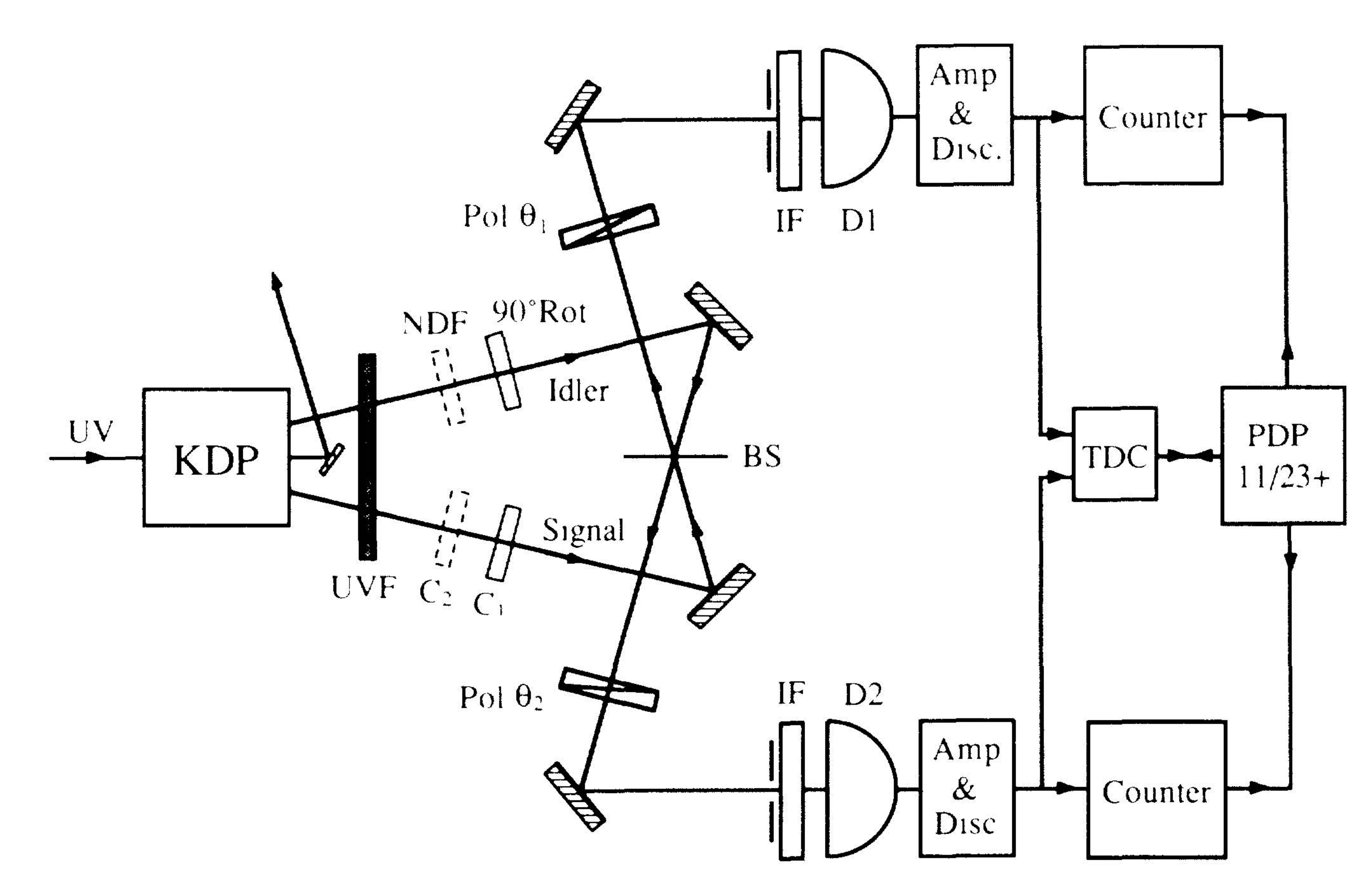}}
\par} \caption{\label{ou1} Outline of the experiment used in~\cite{ou}.}
\end{figure}

The schematic of the experiment in~\cite{ou} (see
also~\cite{2002}) is shown in Fig. \ref{ou1}. In this type of the
experiments, both down-converted photons are split at the beam
splitter (BS), and each detector therefore counts a combination of
both photons.

After the beam splitter, the photons A and B are transformed into
two mixtures. The mixture $\alpha|A\rangle+\gamma|B\rangle$ goes
to the detector~1, and the mixture
$\beta|A\rangle+\delta|B\rangle$ goes to the detector~2. Because
of unitarity of the transformation we must have
$$|\alpha|^2+|\beta|^2=|\gamma|^2+|\delta|^2=1.$$
The state of the two photons is described now by the product wave
function
\begin{equation}\label{doc0}
\psi=[\alpha|A\rangle+\gamma|B\rangle]\otimes
[\beta|A\rangle+\delta|B\rangle],
\end{equation}
where the first factor of the direct product is related to the
detector 1, and the second factor is related to the detector 2.
What is the probability of coincidence counts, if the detectors
efficiency is unity? There are two ways to count this probability.
The first way, and it is commonly accepted, is
\begin{equation}\label{doc}
w_1=|\alpha\delta+\gamma\beta|^2
\end{equation}
and the second way is
\begin{equation}\label{doc1}
w_2=|\alpha\delta|^2+|\gamma\beta|^2.
\end{equation}
Currently, it is assumed that the interference between photons
leads to~$w_1$. On the other hand, the expression for $w_2$ arises
from unitarity or in other words from the conservation law for the
number of particles. Indeed, the probability for both photons to
travel the same path and strike the same detector is
\begin{equation}\label{doc2}
|\alpha\gamma|^2+|\beta\delta|^2.
\end{equation}
The first term of (\ref{doc2}) gives the detection probability of
both photons in the first detector only, and the second term gives
the registration probability of both photons in the second
detector only. According to (\ref{doc1}) the total probability of
counting both photons in a single detector and at a coincidence in
both detectors is
\begin{equation}\label{doc3}
|\alpha\gamma|^2+|\beta\delta|^2+|\alpha\delta|^2+|\gamma\beta|^2=1.
\end{equation}
Thus, the unitarity is not violated. In case of (\ref{doc}) the unitarity is violated, because
\begin{equation}\label{doc4}
|\alpha\gamma|^2+|\beta\delta|^2+|\alpha\delta|^2+|\gamma\beta|^2+
2{\rm Re}(\alpha^*\delta^*\gamma\beta)\neq1.
\end{equation}
The unity is achieved only if the last term in the left-hand side
is equal to zero, which means that there should be no
interference, or no entangled state.

The problem of unitarity is not discussed in the experiments with
the photon splitting, therefore the results of these experiments
cannot be reliably used to prove the action at a distance.

\section{Experiments with neutron interferometer}
\begin{figure}[!h]
{\par\centering\resizebox*{12cm}{!}{\includegraphics{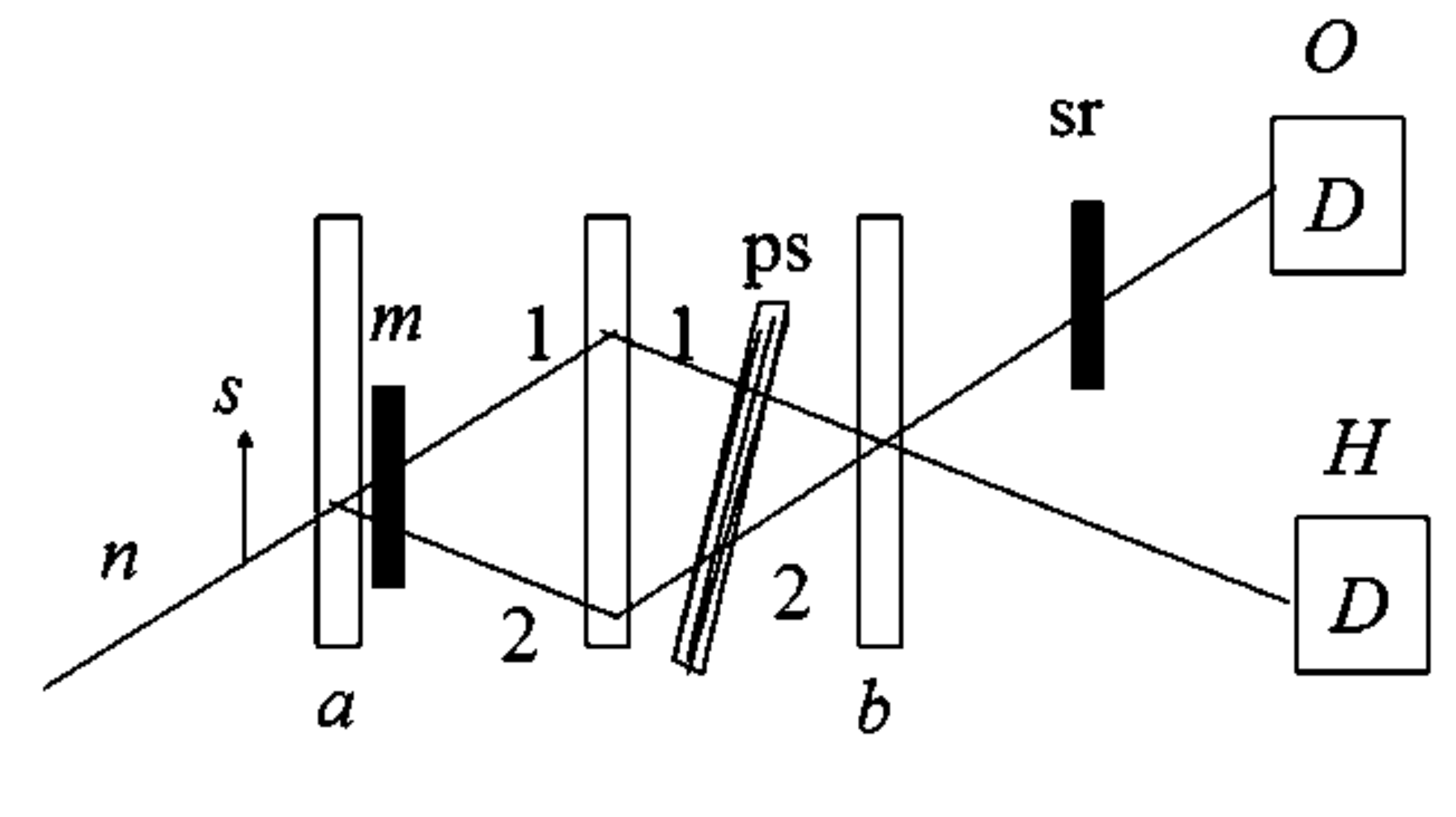}}
\par} \caption{\label{has6a}Generation of entangled state in interferometer}
\end{figure}
The ``disease'' of proving the violation of Bell's inequalities
spreads quickly over science. The experiments with neutron
interferometer~\cite{has1,has2} is a good illustration of this
statement. In~\cite{has1,has2} a neutron interferometer (the
schematic of the experiment is shown in Fig. \ref{has6a}) is used
to create the entangled state for two commuting degrees of freedom
of a single particle
\begin{equation}\label{h1}
|\psi\rangle=A_1|\xi_y\rangle\exp(ikx_1+\varphi_1)\oplus
A_2|\xi_{-y}\rangle\exp(ikx_2+i\varphi_2).
\end{equation}
The first degree of freedom is the spin, and second degree of
freedom is the path that the neutron takes to travel after the
first beam splitter. The first term in (\ref{h1}) describes the
neutron with the spin state $|\xi_y\rangle$ propagating along the
path 1 after the beam splitter, and the second term describes the
same neutron with the spin state $|\xi_{-y}\rangle$ propagating
after the beam splitter along the path 2. To achieve the different
spin states along two paths, a magnet {\bf m} with strong
inhomogeneous field is placed there after the beam splitter $\av$
. According to Stern-Gerlach effect the magnet directs neutrons in
the state $|\xi_y\rangle$ along the path 1, and the neutrons in
the state $|\xi_{-y}\rangle$ along the path 2.

At the beam splitter $b$ the two terms of (\ref{h1}) recombine and
give two new beams, one of which (the beam H) propagates toward
the detector H and another one (the beam O) propagates toward the
detector O. The spin state of the beam O can be represented as
\begin{equation}\label{h2}
|\psi_O\rangle=A\left[|\xi_y\rangle+
r|\xi_{-y}\rangle\exp(i\vartheta)\right],
\end{equation}
where $A$ and $r$ are some constants ($r$ is real), and phase
$\vartheta$ can be varied by the phase shifter {\bf ps} (see fig.
\ref{has6a}). The state (\ref{h2}) is a spinor state
$|\psi_O(\bb)\rangle$ polarized along some direction $\bb$:
\begin{equation}\label{h3}
|\psi_O(\bb)\rangle=C\fr{I+\bb\si}{\sqrt{2(1+b_z)}}|\xi_z\rangle,
\end{equation}
where $C$ is a normalization constant, determined from
\begin{equation}\label{h4}
|C|^2=\langle\psi_O(\bb)||\psi_O(\bb)\rangle=|A|^2\left[\langle\xi_y|+
r\langle\xi_{-y}|\exp(-i\vartheta)\right]\left[|\xi_y\rangle+
r|\xi_{-y}\rangle\exp(i\vartheta)\right]=|A|^2(1+r^2).
\end{equation}
The vector $\bb$ is
$$\bb=\fr1{|C|^2}\langle\psi_O(\bb)|\si|\psi_O(\bb)\rangle=\fr1{1+r^2}\left[\langle\xi_y|+
r\langle\xi_{-y}|\exp(-i\vartheta)\right]\si\left[|\xi_y\rangle+
r|\xi_{-y}\rangle\exp(i\vartheta)\right]=$$
\begin{equation}\label{h5}
=\fr{(2r\sin\vartheta,1-r^2,2r\cos\vartheta)}{1+r^2},
\end{equation}
and $|\bb|=1$.

The experiment~\cite{has1} uses also a spin rotator {\bf sr} (see
Fig. \ref{has6a}), which rotates the spin around $x$-axis to any
desirable angle $\phi$. In addition, an analyzer is placed after
the spin rotator, which transmits neutrons polarized along
$z$-axis. Therefore the count rate of the detector O is
$$I_O=\fr{|C|^2}{4(1+b_z)}
\left|\langle\xi_{z}|\exp(i\phi\sigma_x)(I+\si\bb)|\xi_z\rangle\right|^2=
|C|^2\fr{|(1+b_z)\cos\phi+i(b_x+i b_y)\sin\phi|^2}{4(1+b_z)}=$$
\begin{equation}\label{hag1}
=\fr{|A|^2}4\left[1+r^2+\sqrt{1+r^4+2r^2\cos(2\vartheta)}\cos(2\varphi+\alpha)\right],
\qquad \alpha={\rm
arccot}\left(\fr{2r\cos\vartheta}{1-r^2}\right).
\end{equation}
We see that the intensity in the detector O oscillates
proportionally to $\cos(2\varphi+\alpha)$, where the phase depends
on the phase shift $\vartheta$, and the contrast of the
oscillations\footnote{I regret that in published
version~\cite{conc} there are errors in expressions (49) and
(50)},
\begin{equation}\label{added}
V=\fr{\sqrt{1+r^4+2r^2\cos(2\vartheta)}}{1+r^2},
\end{equation}
also depends on $\vartheta$.

It would be very interesting to check the dependence of $I_O$ on
different parameters, however the goal of~\cite{has1} is
different. It is revealed in the first paragraph of the
paper~\cite{has1} (similar description is presented also
in~\cite{has2}):
\bigskip

\centerline{\parbox{14cm}{\bf ``The concept of quantum
noncontextuality represents a straightforward extension of the
classical view: the result of a particular measurement is
determined independently of previous (or simultaneous)
measurements on any set of mutually commuting observables. Local
theories represent a particular circumstance of noncontextuality,
in that the result is assumed not to depend on measurements made
simultaneously on spatially separated (mutually noninteracting)
systems. In order to test noncontextuality, joint measurements of
commuting observables that are not necessarily separated in space
are required. ''}}
\bigskip

By proving the violation of Bell-like inequalities and the
presence of contextuality the authors want to prove that they deal
with quantum phenomenon. This goal is very surprising, because
such conclusion is very obvious since the interference is really a
quantum phenomenon. The physics becomes like the modern art, which
requires a lot of imagination, but not understanding.

\section{Conclusion}

In this paper I have shown that the non-commutativity of position
and momentum operators does not forbid for particles to have well
defined positions and momenta simultaneously. Therefore the EPR
verdict that QM is not complete is invalid.

I have shown that according to local realism an entangled state of
two particles can be considered as a list of possible product
states realized in every single trial. To check whether such view
is correct, we propose an experiment to measure the ratio of the
transmission probabilities through perpendicular and parallel
analyzers for two photons after the cascade decay. I have shown
that the experiments performed to date do not contradict this
view.

I have analyzed several down-conversion experiments and showed
that they do not prove the non-locality, but that they should be
improved.

I have shown that demonstrating the violation of Bell's
inequalities should not be identified with the proof of QM
non-locality. This is especially clear in the case of experiments
with neutron interferometer, where everything can be explained
from the positions of the local realism.

Finally, I understand that I am embarking on a controversial
topic. It is very important for the reader to abandon his/her
pre-conceived scepticism, backed by the vast history of this
subject, and to understand my arguments in details. Only through
constructive debate can the truth be found. The constructive
criticism is unfortunately absent from the referees reports, whose
main critical points are based on the historical length of the
subject, the number of people backing up existing theory and the
prominence of the original founders (Einstein and others). While
such arguments are valid in presidential elections, they do not
apply to science.

\section{History of submissions, rejections and discussions with referees}

In this article I would like to also include the history of
submissions and rejections of this paper to different journals.
Throughout this somewhat long process the paper underwent a number
of improvements, sometimes according to the the referee reports,
sometimes on my own.  So, this article looks different from the
initial paper submitted to Phys. Rev. A, but the core content
remains the same.

\subsection{Submission to Phys.Rev.A, section ``Fundamental physics''}

On March 12, 2007 I submitted this paper to Phys.Rev.A, the
Fundamental Physics section. On March 13 I received the submission
acknowledgement, and on March 20, only one week after the
submission, the article was rejected directly by the editor. His
letter is attached below.

Re: AC10282

Dear Dr. Ignatovich,

We are sorry to inform you that your manuscript is not considered
suitable for publication in Physical Review A.  A strict criterion
for acceptance in this journal is that manuscripts must convey new
physics.  To demonstrate this fact, existing work on the subject
must be briefly reviewed and the author(s) must indicate in what
way existing theory is insufficient to solve certain specific
problems, then it must be shown how the proposed new theory
resolves the difficulty.  Your paper does not satisfy these
requirements, hence we regret that we cannot accept it for
publication.

Yours sincerely,

Gordon W.F. Drake Editor Physical Review A Email:
pra@ridge.aps.org Fax: 631-591-4141 http://pra.aps.org/

I replied that I do not agree with his decision. The direct
rejection unfortunately does not allow for further discussion, so
I submitted the article to the Russian journal UFN the day I
received the rejection from Phys.Rev. A on March 20.

\subsection{Submission to UFN (Uspekhi)}

On July 24 I received the email reply with two attached letters.
The first letter was the official rejection:

Dear Dr. Ignatovich, your paper and the report of an independent
referee were considered. Taking into account critical character of
the report we decided to refuse from publication of the paper in
UFN. Referee report is attached.

Signed by Deputy Chief Editor, corresponding member of RAS prof.
O.V.Rudenko on June 15 2007.

\subsubsection{Referee report}

The paper criticizes the current approach to analysis of QM
paradoxes, based on EPR effect, Bell's inequalities and Aspect's
experiments. The author claims that the EPR paradox does not
exist, and Aspect's experiments (where it was shown that Bell's
inequalities are violated) prove nothing. The questions raised in
the paper are of interest and they are actively discussed in
literature. However formulation and discussion of these questions
in the submitted article looks not sufficiently complete, and
deduced conclusions are not well based. The main defects of the
author's approach are the following.
\begin{enumerate}
\item The logic of proofs is not very clear, which is very
important for this matter. \item There are no detailed comparisons
of author's conclusions with the current point of view and no
analysis of differences is given. Situation is especially
complicated by the formalism which the author uses for derivation
of specific equations. His formalism is more complicated than the
usual one. So the proofs become non transparent. For instance, the
relation (25), derived by the author contradicts to currently
accepted Eq. (20), however the author does not discuss
sufficiently clear the reasons of such difference. The
nontransparent logic of presentation and absence of legible and
multifold comparisons with conclusions with current ones makes the
author's results dubious. There is one essential defect in the
author's logic, which seems becomes the reason of all the
misunderstandings. He uses a different, unusual definition of the
key notion for the given problem, the notion of the ``state'' in
which the observable has a certain value. The author writes "We
shall show that paradox appears only because of definition of
"precise values" of physical quantities as eigenvalues of their
operators. We shall show that it is necessary to define them as
expectation values, then noncommutativity of operators does not
forbid for Q and p to have precise values simultaneously, and the
EPR paradox disappears". However the current definition of
``certain values'' based on eigen values has an evident ground
(the result of measurement in that case is uniquely predicted),
and in the case of expectation value it is unpredictable. It is
clear that this key notion needs discussion and proof. Their
absence invalidates all the conclusions of the article. Because of
it the submitted paper cannot be published in UFN.
\end{enumerate}

\subsubsection{My reply}

I replied to the editors that the referee report was not
interesting, that my paper was submitted to ZhETF where the
referee reply was much more interesting, and that I will not
continue the discussion with the referee of UFN.

\subsection{Submission to ZhETF (JETP)}

After my submission to UFN, I did not receive any acknowledgement,
therefore I submitted the paper to the Russian journal ZhETF
(JETP) on March 27. The next day I received the acknowledgement,
and on April 2nd I received a letter that my manuscript was
accepted for refereing. The referee report was mailed to me on May
24. At that time my paper did not contain discussion of
down-conversion experiments.

\subsubsection{Referee report (translated from Russian)}
I believe that the paper by Dr. Ignatovich should not be published
in ZhETF for the following reasons.

\begin{enumerate}
\item There are many different publications in the scientific and
popular literature devoted to the studies of the EPR ``paradox''.
It is well known that the resolution of the ``paradox'' was
already given by N.Bohr~\cite{bor},

\item To my opinion, the best methodical description of the EPR paradox
is presented in the book

2. A.Sudbery. Quantum mechanics and physics of elementary
particles (Chapter 5, section 5.3)

In that respect it is surprising that V.K.Ignatovich still thinks
that the state of a subsystem (single particle) after a decay is
described by a wave function instead of the density matrix (see
Eq-s (\ref{J4}) and (8)).

\item The problem of Bell's inequalities, which quantitatively shows
the difference between the classical and quantum
description of the world, is indeed a very exciting problem to any
physicist. To date, numerous experimental and theoretical papers
were devoted to the derivation of these inequalities for a wide
set of physical systems including those with high dimensionality,
e.g. Hilbert space.

I recommend [Dr. Ignatovich] to read the paper by

3. N.V.Evdokimov et all UFN (Russian journal Uspekhi) v.166, p.
91-107 (1997); especially the Introduction, where methodical
aspects of the problem are considered.

\item It appears that the arguments of V.K.Ignatovich in his views
about the experiments on Bell's inequalities are highly biased. My
opinion is based on the fact that [Dr. Ignatovich's] ``proof'' of
the relation (\ref{prr}) is based on the experiments performed
with large uncertainties. See for example the experiments with the
two-photon cascade decay of atoms. Due to the high complexity of
the setup and the measurement technique  at the time, the
measurement precision is not too high.  At the same time, numerous
experiments with the parametric down-conversion of photons were
performed in the past ten years. Relative simplicity of these
experiments and the availability of the efficient photon counters
and driving electronics allowed to significantly increase the
measurement precision. I point out just a few of [these
experiments]~\cite{kwi93,pit,ou,kwi95,kwi99}.

In the modern experiments with quantum optics, the coincidence
counting techniques utilize time windows below 1ns, thus making
the number of the accidental coincidence counts negligibly small.
For example, the number of the accidental coincidences
in~\cite{kwi99} was less than 1Hz, while the true coincidence
count was 1.5 kHz in the maximum of the interference pattern.

One can find a large number of such examples. All these
experiments prove the validity of the orthodox version of the
Bell's inequality. Why wouldn't V.K.Ignatovich apply his
``theory'' [sic] to these experiments?

\item Unfortunately I am not a specialist in the neutron interferometry.
However I can assume that in this case, too, the
adequacy of the author's ``theory'' is based on the experiments
conducted with not high enough precision, thus the presence of the
background allows [the author] to interpreted the results in his
favor. I consider such approach to be one sided. I think it is
inappropriate to publish such arguments, which in addition are
presented in a highly ambitious way (":experiments are shown to
prove nothing"), in a journal of such level as ZhETF.

\end{enumerate}
\subsubsection{My reply to the referee}
\begin{enumerate}
\item Bohr~\cite{bor} in his article did not resolve the paradox.
Instead he proposed
\bigskip

\centerline{\parbox{14cm}{\bf ``a simple experimental arrangement,
comprising a rigid diaphragm with two parallel slits, which are
very narrow compared with their separation, and through each of
which one particle with given initial momentum passes
independently of the other.''}}
\bigskip

In this arrangement either both particles positions or momenta are
measured.
\bigskip

\centerline{\parbox{14cm}{\bf ``In fact to measure the position of
one of the particles can mean nothing else than to establish a
correlation between its behavior and some instrument rigidly fixed
to the support {\Large\it which defines the space frame of reference}.
Under the experimental conditions described such a measurement
will therefore also provide us with the knowledge of the location,
otherwise completely unknown, of the diaphragm with respect to
this space frame when the particles passed through the slits.''}}
\bigskip

From his explanations it follows that after installation of a
diaphragm rigidly fixed to support to measure position of a
particle in my room, I immediately establish the space frame of
reference and geometry in the whole world, and all the other
scientists are now able to measure only positions of their
particles and no one can measure momenta, because it is not known
whether the measured particles were not interacting some time ago.
Shortly: there is nothing in this paper about paradox but
philosophy about complementarity.

\item Presentation of the EPR paradox in the book~\cite{sud} is
the standard one. I found nothing about density matrix in the
section pointed out by referee. The states of free single
particles and entangled states of two particles are always pure
ones. The density matrix appears when one describes a beam of
particles, where averaging over some parameters is necessary. So I
do not see any error in using pure states instead of density
matrix. More over, I found no density matrix in all the papers
describing down conversion experiments, which I had read.

\item With a great pleasure and surprise I had read the
paper~\cite{evd}, and I had found there the following. Let's look
page 93 at the left\footnote{The following text is translated by
me. Readers are advised to look English edition of this paper}
\bigskip

\centerline{\parbox{14cm}{\bf ``There rises a natural question:
how, for instance, position of a switcher at A can affect the
color of lamp, which will be switched on at B?''}}

and further at the right:
\bigskip

\centerline{\parbox{14cm}{\bf``Frequently in the context of
quantum experiments the answer is {\it this is a result of the
quantum non locality}. It means presumably that there is some
mysterious super light influence of an apparatus in A on the event
in B and vice versa.''}}

I have to say that it is just this meaning is supposed in quantum mechanics.
The word {\bf``presumably''} shows that the paper has no relation to the EPR paradox.
The following paragraph on the same page proves it:
\bigskip

\centerline{\parbox{14cm}{\bf``observation of correlations means
transfer of information (protocols of tests with fixation of
current number of trial) from A and B to a third observer C via
ordinary communication channels.''}}

It is great! Therefore when a molecule decays at C into two atoms
with spin 1/2 two parallel analyzers at A and B will not count in
coincidence, because they sent in advance an information to the
molecule, and it will decay to atoms with their spins exactly
along the direction of the analyzers! Therefore the Wheeler's
proposal to deceive the molecule by switching randomly direction
of one of analyzers after the decay is very important! However the
God is great! He knows the maleficence of experimentalists and he
knows construction of generators of random numbers, so he foresees
the outcome of the generator and communicate it to the molecule!
How great becomes the physics. It should be taught in churches,
--- not in the Universities!

\item Thanks to referee for this remark. I thought that
dethronement of the experiments by A.Aspect et all is sufficiently
important, because till now they are considered as a main
experimental proof of the quantum non locality. For example, let's
look at the paper~\cite{hay}. In the introduction we read
\bigskip

\centerline{\parbox{14cm}{\bf ``The concept of entanglement has
been thought to be the heart of quantum mechanics. The seminal
experiment by Aspect et al. \cite{asp1} has proved the "spooky"
nonlocal action of quantum mechanics by observing violation of
Bell inequality with entangled photon pairs.''}}

However the referee is right. It is necessary to review at least
some (there is a flood of them in literature) experiments on
parametric down conversion of photons and to show that they do not
prove the action at a distance. Measurement of violation of Bell's
inequalities, which is equivalent to sum
$$\cos(-45^\circ)+2\cos(45^\circ)-\cos(135^\circ)=2\sqrt2>2$$
is wasting of time and resources, and it has
nothing to do with physics. The experiments themselves are
interesting and genuine and they can be used to study physics of
down conversion, but instead of it the authors are satisfied to
calculate $2\sqrt2$, because it rises their work to the rank of
``basic'' research!
\end{enumerate}

I sent my replies without the new version of the manuscript on
August 1, and received only a reply that my letter was forwarded
to the referee. On September 6 I sent a new version of my
manuscript, which contained the discussions of the down-conversion
experiments. On October 29 I sent an updated version of my
manuscript. On the same day I received the final rejection and the
second referee report. I had no opportunity to continue the
discussion, but I wanted to comment on his report. I added my
comments in italic below.

\subsubsection{The second referee report}

The author discusses the foundations of quantum mechanics.
Specifically, he analyzes the famous Einstein-Podolsky-Rosen
paradox in the original and Bohm formulations, discusses briefly
the Bell inequalities. He suggests new formulations of basic
principles of quantum mechanics, which in his view resolve the
problems, and analyzes some experiments in order to demonstrate
that they in fact do not support the standard view but rather are
inconclusive or support his view. I do not find the paper suitable
for publication in JETP since the problems discussed are not
considered as such any more; furthermore, the author suggests to
replace the basic concepts of quantum mechanics to ``resolve'' the
problems --- this can hardly be viewed as a resolution. I discuss
this in more detail below. Further, the author does not properly
describe the context of this work and does not define precisely,
which physics problems are being solved; this makes the manuscript
hard to follow.

{\it The referee basically says that the debate is closed for the
EPR paradox, thus no more articles should be published. I am
somewhat at a loss on how to comment on this statement, except
that I disagree with the referee's opinion. }

Through the paper he stresses and ``illustrates'' that the
uncertainty principle is invalid and does not make sense.

{\it This argument is not true. I claim that the uncertainty
principle is valid, but it does not have the crucial role that it
is given in quantum mechanics. }

Instead, he suggests that, for instance, the coordinate and
momentum can simultaneously have well definite values (no
fluctuation)

{\it There are indeed no fluctuations, but there is dispersion.
For example, the Gaussian wave packet has precisely defined
position, but it also has width, which is the dispersion. It was
von Neumann, who proved that there are no dispersion free states
for position and momentum operators.}

To achieve this, he proposes (in the context of the work by
Einstein, Podolsky and Rosen) in order ``to avoid paradox'', to
``reject the definition'' of the concept of ``an operator A having
some value $a$ with certainty (without fluctuations)'': while this
happens only when the system is in eigenstate of A with the
eigenvalue $a$, he suggests that the operator always has a certain
value $\langle\psi|A|\psi\rangle$, which we usually refer to as
the average value. In this sense, he throws away a significant
notion of the uncertainty principle, and merely observes that
every operator has some average value in any state. Similarly, he
``shows'' that all components of a spin-half can simultaneously
have well defined values, without fluctuations; this is again
achieved by replacing the notion of an eigenvalue by that of an
average. He disregards the fact that the fluctuations,
$\langle(A-\langle A\rangle)^2\rangle$, are vanishing only in
eigenstates of A.

{\it If this referee happened to review the book of von Neumann,
the book would never got accepted for publication. As for spin
half particle: suppose that the particle is polarized along a
vector $\av$. The average (expectation value in my opinion) of the
Pauli matrix (operator) $\sigma_x$ is $a_x$. What referee calls
fluctuation is
$$\langle(\sigma_x^2-a_x^2)\rangle\equiv\langle(1-a_x^2)\rangle\equiv1-a_x^2=
a_y^2+a_z^2.$$ It is a fluctuation by definition, but not by
physics. It means that the actual value of the component $a_x$
depends on the choice of the coordinate axis $x$ and nothing else.
Moreover, the operator $1-a_x^2$ is just an operator, for which
the value can also be found as an expectation value. }

I should stress that the ''EPR paradox'', and its reformulation
due to Bohm, are not considered as paradoxes, but are used in the
modern literature to illustrate the non-local nature of quantum
mechanics  and the notion of entanglement, as well as to show how
quantum mechanics contradicts the classical intuition. They also
demonstrate that, unlike the classical physics, the measurement
(of one particle of the two) influence the ``state'' of the other
particle (or rather the result of subsequent measurements on this
other particle) provided the particles are in an entangled state.
In this sense it is impossible to say that, say, one spin in a
singlet pair has a well defined direction, which one could try to
``guess'' (as the author suggests).

{\it The non-local nature of quantum mechanics is direct
consequence of the paradox. The action at a distance is introduced
to resolve the paradox and to avoid conclusion that QM is not a
complete theory. But if there was no paradox to start with, the
notion of action at a distance would never appear. Resolving the
paradox in an alternative way casts doubt on the accepted notion
of entanglement. While QM challenges the classical intuition, it
should not be the source of miracles. We of course can debate the
issue for very long time, but why not just attempt to measure the
coincidence counts of two photons in Aspect's a.o. experiments
with two perpendicular analyzers?}

In the discussion and derivations the author implicitly (or
explicitly) replaces the rules of quantum mechanics to derive
``new results'' or `` resolve problems''. For example, in eqs.
(\ref{J9}) and (\ref{n2e}) he adds probabilities rather than
amplitudes, which one can also describe by saying that he assumes
that the state (the spin singlet or the state of two photons after
decay) is a mixture, rather than a superposition, of ``classical''
states, where in each constituent state in the superposition one
spin has direction $\OO$, and the other --- $-\OO$ ($\OO$ being an
arbitrary unit vector). Using these ``rules'' he derives new
expressions for the probabilities of detection.

{\it This statement is again not true. I do not use the mixture of
states. I use pure product states, but I average the final
probabilities obtained with these pure states over the initial
states after decay. I basically follow what we always do in
scattering theory. We obtain scattering amplitude, square it
modulo then sum over the final states and average over the initial
ones. It does not mean that we use mixtures.}

\section*{Acknowledgement}
I greatly appreciate very interesting discussions with my son
F.V.Ignatovich and his advices. They helped me very much to
improve the paper. I also very grateful to editorial board of
Concepts of Physics, and especially to Executive Editor Edward
Kapuscik for constructive attitude to my works.


\begin{thebibliography}{99}
\bibitem{epr}
A. Einstein, B. Podolsky, and N. Rosen. Physical Review. May 15,
1935, v47, pp 777-780.
\bibitem{ahb}
D.Bohm and Y.Aharonov.``Discussion of experimental proof for the
paradox Einstein, Rosen, and Podolsky,'' Phys. Rev. 108 (1957)
1070.
\bibitem{bel}J.S.Bell. Physics 1 (1964) 195; J.S.Bell. ``Speakable
and unspeakable in quantum mechanics.'' Cambridge University
press, 2004, p. 14.
\bibitem{hay}
Masahito Hayashi et all, Hypothesis testing for an entangled state
produced by spontaneous parametric down-conversion, Phys. Rev. A
74, 062321 (2006)
\bibitem{asp1}
A.Aspect, P.Grangier, G.Roger, Experimental tests of realistic
local theories via Bell's theorem.  Phys. Rev. Let. V.47, (1981)
P. 460.
\bibitem{koco}
C.A.Kocher, E.D.Commins, ``Polarization correlation of photons
emitted in an atomic cascade.'' PRL 18 (1967) 575-577.
\bibitem{asp2}A.Aspect, P.Grangier, G.Roger, Experimental
Realisation of Einstein-Podolskii-Rosen-Bohm {\it
Gedankenexperiment}: A new violation of Bell's inequalities. Phys.
Rev. Let. V. 49 (1982) 91.
\bibitem{asp3}
A.Aspect, J.Dalibard, G.Roger, Experimental test of Bell's
inequalities, using time-varying analyzers. Phys. Rev. Let. V. 49
(1982) P. 1804.
\bibitem{has1}
Y.Hasegawa, R.Loidi, G.Badurek, M.Baron, H.Rauch,{\it Violation of
a Bell-like inequality in single-neutron interferometry.} Nature,
V.425 (2003) P. 45-48.
\bibitem{has2}
Y.Hasegawa, R.Loidi, G.Badurek, M.Baron, H.Rauch,{\it Quantum
contextuality in neutron interferometer experiment.} Physica B V.
385-386, Part II, (11 November 2006) P. 1377.
\bibitem{neum}
J. von Neumann, {\it Mathematical Foundations of Quantum
Mechanics.} (Princeton University Press, Princrton, New Jersey,
1955) Ch.IV, sec.1 and 2.
\bibitem{neum1} J.Albertson, Am. J. Phys. 29 (1961) 478.
\bibitem{conc}
V.K.Ignatovich "On uncertainty relations and interference in
quantum and classical mechanics" Concepts of Physics, V..III,  p.
11, 2006.
\bibitem{kwi99}
P.G.Kwiat, E.Waks, A.G.White, I.Appelbaum, and P.H.Eberhard,
Ultrabright Source of Polarization Entangled Photons. Phys.Rev.A,
60, R773-R776 (1999).
\bibitem{kwi95}
7. P.G.Kwiat, K.Mattle, H.Weinfurter, A.Zeilinger, A.V.Sergienko,
Y.H.Shih, New High-Intensity Source of Polarization-Entangled
Photon Pairs. Phys.Rev.Lett., 75, 4337-4341 (1995).
\bibitem{zei}
P.G.Kwiat et all, Phys.Rev.A, 66, 013801 (2002).
\bibitem{kwi93}
P.G.Kwiat, A. M. Steinberg, and R. Y. Chiao, ``High-Visibility
Interference in a Bell-Inequality Experiment for Energy and
Time.'' Phys. Rev. A, {\bf 47}, R2472-R2475 (1993).
\bibitem{pit}
T.B.Pittman, Y.H.Shih, A.V.Sergienko, and M.H.Rubin,
``Experimental Test of Bell's Inequalities Based on Spin and
Space-Time Variables.'' Phys.Rev.A, {\bf 51}, 3495-3498 (1995).
\bibitem{ou}
Z.Y.Ou and L.Mandel, Violation of Bell's Inequality and Classical
Probability in a Two-Photon Correlation Experiment.
Phys.Rev.Lett., 61, 50-53 (1994).
\bibitem{bor}
N.Bohr, Can Quantum Mechanical Description of Physical Reality Be
Considered Complete? Phys.Rev., 48, 696-702 (1935). We develop and
\bibitem{sud}
A.Sudbery. Quantum Mechanics and the Particles of Nature: An Outline for Mathematicians.
Cambridge University Press, London, 1986.
\bibitem{evd}
N.V.Evdokimov et all UFN (Russian journal Uspekhi) v.166, p.
91-107 (1997)
\bibitem{2002}
Giovanni Di Giuseppe, Mete Atat\"ure, Matthew D. Shaw, Alexander
V. Sergienko, Bahaa E. A. Saleh, and Malvin C. Teich,
``Entangled-photon generation from parametric down-conversion in
media with inhomogeneous nonlinearity.'' Phys. Rev. A {\bf 66},
013801 (2002).
\bibitem{conc}
V.K.Ignatovich, ON EPR PARADOX, BELL'S INEQUALITIES AND
EXPERIMENTS THAT PROVE NOTHING, Concepts of Physics, the old and
new, v. 5, No 2, pp. 227-272, 2008.
\end{thebibliography}
\end{document}